\documentclass[preprint,12pt]{elsarticle}

\usepackage{tcolorbox}
\usepackage{amssymb}
\usepackage{url} 
\usepackage{float}

\usepackage{longtable} 
\usepackage{hyperref} 
\usepackage{tabularx} 
\usepackage{svg}
\usepackage{booktabs}
\usepackage{array}
\usepackage[utf8]{inputenc}
\usepackage{longtable}
\usepackage{multirow}
\usepackage{lmodern}
\usepackage{lipsum} 
\usepackage{graphicx} 
\usepackage{adjustbox}
\usepackage{xcolor}
\usepackage{tcolorbox}
\usepackage{pifont} 
\usepackage{pgfplots}
\pgfplotsset{compat=1.18} 
\usepackage[utf8]{inputenc}
\usepackage{comment}
\usepackage{pgfplots}
\pgfplotsset{compat=1.18}

\usepackage{pgfplotstable}

\usepackage{xcolor} 
\usepackage{soul}
\usepackage[T1]{fontenc}

\usepackage{multirow}
\usepackage{tablefootnote}
\usepackage{threeparttable}
\usepackage{enumitem}

\journal{}
\DeclareUnicodeCharacter{FF08}{(} 
\DeclareUnicodeCharacter{FF09}{)} 
\begin{document}
\renewcommand{\arraystretch}{1.2} 
\setlength{\LTcapwidth}{\textwidth} 

\begin{frontmatter}



\title{Large Language Models (LLMs) \\for Requirements Engineering (RE): \\A Systematic Literature Review}


\author[inst1]{Mohammad Amin Zadenoori\footnote{\url{amin.zadenoori@unipd.it}}}
\affiliation[inst1]{organization={Department of Statistics, University of Padova},
            addressline={Via Cesare Battisti 241},
            city={Padova},
            postcode={35121},
            country={Italy}}

\author[inst2]{Jacek Dąbrowski\footnote{\url{jacek.dabrowski@lero.ie}}}
\affiliation[inst2]{organization={Lero, the Research Ireland Centre for Software, University of Limerick},
            city={Limerick},
            postcode={V94 T9PX},
            country={Ireland}}

\author[inst3]{Waad Alhoshan\footnote{\url{wmaboud@imamu.edu.sa}}}
\affiliation[inst3]{organization={Imam Mohammad Ibn Saud Islamic University (IMSIU)},
            city={Riyadh},
            postcode={13317},
            country={Saudi Arabia}}

\author[inst4]{Liping Zhao\footnote{\url{liping.zhao@manchester.ac.uk}}}
\affiliation[inst4]{organization={University of Manchester},
            addressline={Oxford Road},
            city={Manchester},
            postcode={M13 9PL},
            country={United Kingdom}}

\author[inst5]{Alessio Ferrari\footnote{\url{alessio.ferrari@ucd.ie}}}
\affiliation[inst5]{organization={University College Dublin (UCD)},
            addressline={Belfield, Dublin 4},
            city={Dublin},
            postcode={D04 V1W8},
            country={Ireland}}
\begin{abstract}
Large Language Models (LLMs) are finding applications in numerous domains, and Requirements Engineering (RE) is increasingly benefiting from their capabilities to assist with complex, language-intensive tasks. 
This paper presents a systematic literature review of 74 primary studies published between 2023 and 2024, examining how LLMs are being applied in RE. The study categorizes the literature according to several dimensions, including publication trends, supported RE activities, prompting strategies, and evaluation methods. 
Our findings indicate notable patterns, among which we observe substantial differences compared to previous works leveraging standard Natural Language Processing (NLP) techniques. Most of the studies focus on using LLMs for requirements elicitation and validation, rather than defect detection and classification, which were dominant in the past. Researchers have also broadened their focus and addressed novel tasks, e.g., test generation, exploring the integration of RE with other software engineering (SE) disciplines.  Although requirements specifications remain the primary focus, other artifacts are increasingly considered, including issues from issue tracking systems, regulations, and technical manuals. The studies mostly rely on GPT-based models, and often use Zero-shot or Few-shot prompting. They are usually evaluated in controlled environments, with limited use in industry settings and limited integration in complex workflows. Our study outlines important future directions, such as leveraging the potential to expand the influence of RE in SE, exploring less-studied tasks, improving prompting methods, and testing in real-world environments, considering the usefulness of these tools, besides their performance. Our contribution also helps researchers and practitioners use LLMs more effectively in RE, by providing a list of identified tools leveraging LLMs for RE, as well as datasets.

\end{abstract}




\end{frontmatter}


\section{Introduction}


The central role of Natural Language Processing (NLP) in Requirements Engineering (RE) has been consistently studied \cite{vanLamsweerde2000, Sommerville2016}. Early observations, such as Abbott and Moorhead's \cite{Abbott1981} assertion that ``natural language is the most suitable language for requirements'', laid the groundwork for its adoption in RE. However, traditional NLP approaches—--those preceding the advent of deep learning, such as rule-based systems and statistical methods—--often struggled to handle the nuances, ambiguities, and complexities inherent in real-world requirements documents. These limitations stemmed from their reliance on predefined rules or shallow statistical models, which lacked the ability to deeply contextualize language.


The introduction of deep learning marked a shift in NLP capabilities, enabling more robust handling of linguistic intricacies through neural network architectures and word embeddings~\cite{ferrari2018identification}. This evolution progressed further with the development of transformer-based models, exemplified by BERT (Bidirectional Encoder Representations from Transformers) \cite{devlin2019bertpretrainingdeepbidirectional}, which introduced bidirectional context understanding, enhancing tasks like requirements classification \cite{10.1007/978-3-031-88531-0_13} and traceability \cite{lin2021traceability} in RE. Generative Large Language Models (LLMs) made another step forward. These models are designed to create new content, such as text, by predicting and generating sequences of words based on vast training data. Prominent examples of LLMs include OpenAI's GPT models (e.g., ChatGPT \cite{OpenAI2023}), Meta's LLaMA \cite{Touvron2023}, and DeepSeek-R1 \cite{deepseekai2025deepseekr1incentivizingreasoningcapability}. While BERT, a bidirectional-only encoder, excels at understanding context for tasks like classification or question-answering, generative LLMs use autoregressive or similar architectures to produce coherent, contextually relevant outputs, often for creative writing, dialogue, or code generation. Their core difference lies in their purpose: BERT analyzes and interprets, while generative LLMs synthesize and create, leveraging patterns learned from diverse datasets to mimic human-like text production~\cite{10.5555/3495724.3495883}. 
Powered by Deep learning and expansive transformer architectures, these LLMs offer transformative potential for RE, and their use in RE research has quickly become widespread.

This Systematic Literature Review (SLR) investigates the research landscape of LLMs for RE (LLM4RE), i.e., the use of LLM to support RE-relevant activities. We focus our study on generative pre-trained transformers---commonly referred to as LLMs and their successors \cite{see2019massivelypretrainedlanguagemodels}---rather than encoder-only transformers, such as BERT.  The reason for this choice is that we want to focus on models whose capabilities extend beyond text representation and classification, and encompass advanced abilities in text understanding and production. Furthermore, the introduction of LLMs has marked a paradigm shift in everyday usage of NLP techniques, and we want to see how the RE field is being shaped by these revolutionary tools.



In this work, we identified 74 primary studies published mostly in conferences (66\%) and workshops (23\%), reflecting a collaborative, early-stage research landscape. The analysis focuses on core RE tasks such as \textit{Requirements Elicitation} and \textit{Validation}. Prompt engineering strategies---particularly \textit{Zero-shot} (44\%) and \textit{Few-shot} (29\%) prompting---dominate the field, while emerging techniques like \textit{RAG} (6\%) and \textit{Interactive prompting} (5\%) remain underexplored. \textit{GPT-based models} lead adoption (90\%), accompanied by a smaller but growing interest in alternatives such as \textit{LLaMA} (15\%). Many studies (61\%) share their source code and prompts, indicating a move toward open science, but shared datasets are still scarce. Evaluation still relies heavily on \textit{Laboratory Experiments} (75\%), which highlights the need for real-world validations and field studies. Underexplored areas (e.g., \textit{Requirements Retrieval}, \textit{Terminology Extraction}), more diverse models, advanced prompting methods, as well as deeper practitioner collaboration, present key opportunities for further advancing LLM4RE, and broaden its impact on the Software Engineering (SE) field and beyond.

The remainder of the paper is structured as follows: Section 2 presents the concepts and definitions. Section 3 discusses related works. In Section 4, we describe the research methodology. Section 5 covers the results, while Section 6 provides a summary and discussion. Section 7 outlines the threats to validity, and the final section presents the conclusions.

\textbf{Data Availability.}
The spreadsheets resulting from our data extraction and grouping, along with all synthesized data, including generated statistics and classifications, are publicly available in the supplementary material of this survey \cite{amin_zadenoori_2025_17068810}.

\textbf{Disclaimer.} This paper is a preliminary study. No systematic cross-checking has been performed on the retrieved and analysed studies. The paper will be updated, and future versions will include appropriate measures to increase the reliability of the findings.
\section{Background}



This section introduces key concepts associated with RE and LLMs. 
Given these premises, we then outline the different dimensions investigated in our study, which are the foundations of the classification scheme used in this paper to categorise the research contributions.

\subsection{Requirement Engineering}

Requirements Engineering (RE) is the subfield of Software Engineering (SE) concerned with the systematic processes of eliciting, documenting, analyzing, verifying, validating, and managing system and software requirements throughout the entire system lifecycle~\cite{ferrari2025handbook}. RE activities are predominantly conducted in natural language (NL), which serves as the most widely understood medium of communication. Using NL facilitates effective knowledge exchange among the diverse stakeholders involved in system development, including customers, developers, business analysts, and assessors.

RE is structured around a series of interrelated phases, each addressing specific aspects of requirement development and management. The first phase, \textit{Elicitation}, focuses on understanding the goals, objectives, and needs of stakeholders for building a proposed software system. This phase involves gathering information from customers, users, and other stakeholders through interviews, workshops, surveys, and observation, ensuring that the system’s intended functionality and qualities are captured comprehensively. Following elicitation, \textit{Analysis} evaluates the quality of the recorded requirements and identifies potential anomalies, such as ambiguity, inconsistency, or incompleteness. This step ensures that requirements are clear, correct, and feasible before they are used in subsequent activities. Once requirements have been analyzed, the \textit{Modeling} phase translates them into interpretable and structured conceptual representations. These models---ranging from UML diagrams to ontologies---support both technical and non-technical stakeholders in understanding system functionality. The \textit{Management} phase then addresses the organization of requirements over time, handling changes, maintaining traceability between requirements and downstream artifacts, and ensuring alignment with evolving stakeholder needs and project goals. \textit{Validation and Verification (V\&V)} follows, ensuring that requirements accurately reflect stakeholder intent (\textit{Validation}) and that system specifications meet these requirements (\textit{Verification}). In a waterfall-like, idealised software lifecycle, these activities are followed by other SE phases, which heavily depend on requirements, such as \textit{Architecture and Design}, \textit{Implementation}, and \textit{Testing}. 

RE phases typically consist of one or more NL-intensive tasks. These include \textit{Classification}, i.e., categorising requirements into different classes such as functional vs quality, \textit{Tracing}, i.e., defining trace-links between requirements and other artifacts, or \textit{Defect Detection}, i.e., identifying quality issues in requirements. Traditionally, RE tasks revolve around the artifact known as the \textit{System/Software Requirements Specification}, which provides a structured and comprehensive account of the system’s intended functionality, quality attributes, and constraints. The SRS serves as a contractual reference among stakeholders, guiding subsequent phases of software development, from design to testing and maintenance. Its role is twofold: on the one hand, it ensures that stakeholder needs are captured, agreed upon, and communicated in a precise manner; on the other hand, it provides a stable foundation against which system implementation and verification can be conducted. However, in recent years, a broader range of artifacts has entered the RE landscape, complementing the traditional reliance on specifications. These include \textit{User Stories}, which capture functional needs in lightweight, scenario-driven forms; \textit{User Feedback} collected from social media platforms and app stores, offering direct insights into user satisfaction and pain points; \textit{Legal and Regulatory Documents}, which introduce compliance constraints; and other forms of unstructured or complex artifacts. Together, these sources expand the scope of RE beyond classical specification, but also introduce new challenges of interpretation, integration, and consistency management.

\subsection{LLMs}
LLMs represent a class of sophisticated artificial intelligence (AI) systems designed to handle and generate human language in a wide range of contexts. They are trained on extensive corpora of text—often encompassing diverse sources such as books, websites, academic papers, and social media—to learn patterns, structures, and relationships within language. By internalizing these patterns, LLMs can effectively capture semantic nuances, syntactic rules, and contextual cues that allow them to perform various NLP tasks. Among the most prominent examples of LLMs are models like GPT \cite{brown2020language}, which has demonstrated remarkable capabilities in tasks such as text completion, idea generation, and conversational interaction.

These models typically rely on deep learning architectures known as transformers \cite{vaswani2017attention}, a key innovation that introduced the concept of self-attention. Self-attention mechanisms enable LLMs to assign different levels of importance to words (or tokens) in a given sequence, refining the way they capture contextual information. This ability is crucial for maintaining coherence across longer pieces of text, making models adept at tasks like summarization, translation, and question answering. Furthermore, LLMs can adapt to a variety of domains—ranging from legal and medical texts to creative writing—by leveraging transfer learning, where a model pre-trained on extensive general data is fine-tuned on more specialized or smaller datasets.

A notable characteristic of LLMs is their ability to generate coherent and contextually relevant text outputs, even in scenarios that require nuanced understanding. Their scalability allows for handling vast amounts of data, thereby improving their predictions and reducing errors related to grammar, style, and factual consistency (though ongoing research continues to address challenges such as model bias and misinformation). As a result, LLMs have become essential tools in the modern NLP landscape, powering a variety of real-world applications: from chatbots and virtual assistants to intelligent tutoring systems, content moderation, and advanced research support. Given their versatility and potential for automation, LLMs are poised to revolutionize how we interact with and harness the power of language in technology-driven environments.

\subsection{Scope of the Study: Conceptual Scheme}
\label{sec:conceptualmodel}

\begin{figure}[h!] 
    \centering
    \includegraphics[width=1\textwidth]{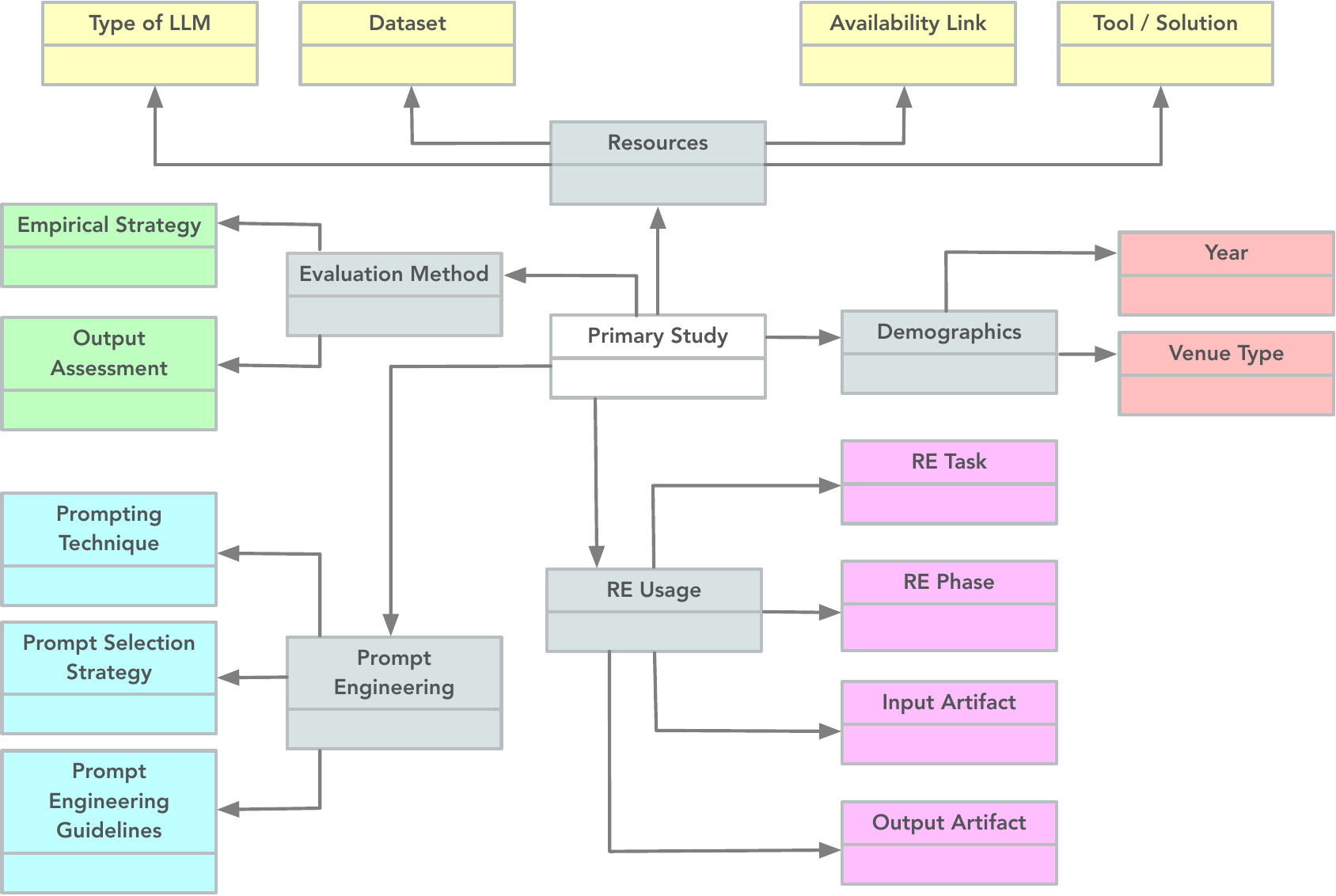} 
    \caption{Conceptual scheme of LLM for RE used as the basis for data extraction.}
    \label{fig:conceptualmodel}
\end{figure}

Fig.\ref{fig:conceptualmodel} represents the main concepts and relations thereof used in this SLR, and that form the basis for our data extraction schemes. The detailed schemes are presented in each section of the data analysis part of the paper, to facilitate interpretation of the findings.

The central unit of analysis is the \textbf{Primary Study}, i.e., the empirical study that analyses the use of LLMs in RE. The primary study has \textbf{Demographics}, including in particular Year and Venue Type, besides other typical information (authors, DOI, etc.)---not represented in the figure. 

A main element of interest is the \textbf{RE Usage}, i.e.,  how the study uses LLMs in RE. This entails addressing a specific \textbf{RE Task}, for example, requirements classification or defect detection, which can be applied during a certain \textbf{RE Phase}, e.g., elicitation, analysis, modelling. The task typically take a main \textbf{Input Artifact}, e.g., a software requirements specification, and produces an \textbf{Output Artifact}, e.g., a graphical model\footnote{More than one artifact can be taken as input, and more than one can be produced as output. Here, for simplicity, we consider the most relevant artifact considered in the primary studies}. Note that the same RE task can be applied in different phases and with different artifacts. For example, the classification task can be applied during the RE phase of elicitation, for example using app reviews as input, or during the management phase, using software requirmeents specification as input.  

To trigger the generative behaviour of the LLM, the study applies a certain \textbf{Prompt Engineering} approach. This consists of \textbf{Prompting Techniques}, e.g., zero-shot, few-shot---typically more than one is used in a paper. Furthermore, it includes a strategy to select the most appropriate phrasing to be used for the prompt (\textbf{Prompt Selection Strategy}), and it can include a reference to the \textbf{Prompt Engineering Guidelines} adopted. 

The study also uses a set of \textbf{Resources}. These are characterised by the \textbf{Type of LLM} adopted, often more than one, and the \textbf{Dataset} used for the empirical evaluation. Furthermore, the tool can make available through a link (\textbf{Avaliability Link}) the source code used and the prompts. Finally, it could incorporate the LLMs in a more complex \textbf{Solution}, e.g., combining both LLMs and standard NLP techniques to perform a certain RE Task, or an integrated \textbf{Tool} that implements the full solution\footnote{Often, in the surveyed papers, it is unclear whether a certain solution has been actually implemented in an integrated tool. For this reason, in this paper, we treat the two types of resources together.}.

A primary study uses  \textbf{Evaluation Methods}. This includes the \textbf{Empirical Strategy}~\cite{stol2018abc} for the design of the study, for example, a laboratory experiment comparing different LLMs, or an experimental simulation, comparing users' performance on a certain task with and without LLMs. The output generated by the LLM is typically assessed (\textbf{Output Assessment}), and this can be done, for example through quantitative methods, e.g., applying precision and recall, or through qualitative ones, for example through thematic analysis~\cite{terry2017thematic} of the generated output. 
\section{Related Reviews}

The integration of LLMs into the field of SE  has shown transformative potential across various tasks. A foundational survey by \cite{10449667} highlights LLMs' emergent properties, such as novelty and creativity, enabling their use in coding, design, repair, refactoring, performance improvement, documentation, and analytics. It emphasizes challenges like mitigating incorrect outputs and advocates for hybrid approaches combining LLMs with traditional SE methods to ensure reliability. Still encompassing the whole SE field, Hou \textit{et al.}\cite{Hou2024} conducted an SLR of 395 publications from 2017 to 2024, categorizing LLM applications in SE, examining data collection, optimization, and evaluation strategies, and identifying supported tasks, while noting gaps for future research. Jin et al. \cite{jin2024llmsllmbasedagentssoftware} go beyond the study of LLMs in isolation and consider LLM-based agents, that is, systems where LLMs are enriched with tools, memory, and decision-making mechanisms to perform more complex and autonomous tasks. Additionally, He \textit{et al. }\cite{10.1145/3712003} explores LLMs in multi-agent systems for SE, conducting a systematic review of their applications across the applications, with case studies illustrating the frameworks’ capabilities and limitations, and proposing a research agenda to enhance agent synergy and autonomy for Software Engineering 2.0.

Some studies have also focused on specific subfields of SE. In software testing, Wang \textit{et al.} \cite{10440574} analyzed 102 studies, showing LLMs’ roles in test case preparation and program repair, discussing prompt engineering, scalability challenges, and innovation opportunities.  LLMs also have an impact on software architecture, automating and enhancing tasks such as design decision classification, pattern detection, and architecture generation. A systematic review \cite{schmid2025softwarearchitecturemeetsllms} analyzed 18 articles and found that LLMs excel in these areas but are underutilized in tasks like code generation and architecture conformance checking, highlighting opportunities for future research to expand their application in software architecture.

Overall, these studies demonstrate the growing interest in applying LLMs to software engineering. However, none of them specifically target the RE field. An extensive literature review of 404 studies was conducted by Zhao \textit{et al.} \cite{10.1145/3444689}, covering research up to 2019 that applied NLP techniques to RE. Our work can be seen as a natural continuation of that effort, tracing the evolution from traditional NLP approaches to the emerging use of LLMs in RE.

\section{Research Methodology}
\label{sec:methodology}
Our study followed the guidelines proposed by Kitchenham \cite{Kitchenham2011}. We first defined the research questions based on the overall research goal. Next, we drafted a review protocol to guide data collection from primary studies. We then performed a literature search and selected studies using inclusion and exclusion criteria. After that, we read the selected studies, extracted relevant data items, and recorded them in data extraction forms. Finally, we consolidated the information and reported it in our study.


\subsection{Research Questions}
\label{subsec:Research Questions}

The overall goal of our SLR is \textit{to provide a broad overview of the application of LLMs for RE} by categorising and summarising the existing literature. Based on this goal, we formulated the following research questions:

\begin{itemize}
\item \textit{RQ1:} What are the demographics of publications on LLMs for RE?

\item \textit{RQ2:} How are LLMs used in RE?
\begin{itemize}
	\item \textit{RQ2.1:} What RE tasks do LLMs support?
	\item \textit{RQ2.2:} What RE phases do LLMs address?
	\item \textit{RQ2.3:} What input artifacts are used?
	\item \textit{RQ2.4:} What output artifacts are produced?
	\item \textit{RQ2.5:} What is the co-occurrence of input and output artifacts?
\end{itemize}

\item \textit{RQ3:} How is prompt engineering applied in LLM-based RE approaches?
\begin{itemize}
	\item \textit{RQ3.1:} What prompt engineering techniques are employed?
	\item \textit{RQ3.2:} What prompt selection strategies are used?
	\item \textit{RQ3.3:} What types of prompt engineering guidelines are followed?
\end{itemize}

\item \textit{RQ4:} What resources are used in LLM-based RE approaches?
\begin{itemize}
	\item \textit{RQ4.1:} What datasets are used?
	\item \textit{RQ4.2:} Are source code and prompts publicly available?
	\item \textit{RQ4.3:} What types of LLMs are used
\item \textit{RQ4.4:} Is a novel tool or solution incorporating LLMs proposed?
\end{itemize}

\item \textit{RQ5:} How are LLM-based RE approaches evaluated?
\begin{itemize}
	\item \textit{RQ5.1:}  How are the outputs of the LLMs assessed?
	\item \textit{RQ5.2:} What primary empirical evaluation strategies are used?

\end{itemize}

\end{itemize}

In answer to RQ1, we report the number of studies published over time and characterize their types and the venues where they have been published. We also highlight the most frequent venues where scholars have published their work, listing key contributions in this research area. With RQ2, we aim to understand the application of LLMs in RE: which tasks and phases they support, what types of input artifacts are provided to them, what output artifacts they generate, as well as their co-occurrence. RQ3 investigates how prompt engineering is applied in LLM-based RE applications, including techniques used, prompt selection strategies, and guideline types reported in the primary studies. In RQ4, we explore availability of supplementary materials of these studies to support  of future replication and extensions studies. More specifically the questios explore the availabilty of used dataset, source code, promtps, LLM types as well as information about using the novel tooling for LLM-based RE approaches. Finally, RQ5 explores how LLM-based RE approaches are evaluated in the literature, focusing on the empirical strategies used and how LLM outputs are assessed.

\subsection{Literature Search and Selection}

\begin{figure}[H]
	\centering
	\includegraphics[width=1\textwidth]{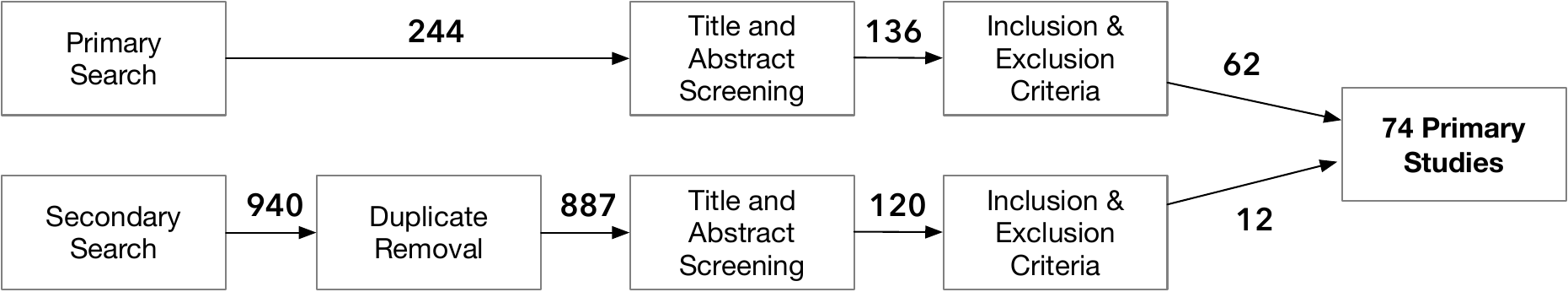}
	\caption{Systematic search process: papers retrieved from the primary and secondary search and filtered through inclusion and exclusion criteria.}
	\label{fig:searchDiagram}
\end{figure}

We followed a systematic literature search and selection strategy to identify relevant studies focusing on the application of LLMs to RE. The process was conducted in two main stages: an initial digital library search and a complementary issue-to-issue search.  Figure~\ref{fig:searchDiagram} summarises the main steps of this process and their outcomes. All steps were performed by the first author of the paper, who double-checked with the last author in case of doubts\footnote{We would like to remark that this is a preliminary version, and in a further, more extensive  publication we will perform cross-checks between  multiple authors to increase the reliability of the findings.}.

\textbf{Primary Search in Digital Libraries.} The primary search was conducted using \textit{Scopus}\footnote{\url{https://www.scopus.com}} digital library, chosen for its comprehensive coverage of high-quality journals and conference proceedings from reputable publishers such as Elsevier, IEEE, and ACM. We formulated a search query that we applied to both the meta-data and full-text (when available) of the publication. The query was formulated using major terms (and their synonyms) referred to in our research goal. The adopted string is as follows.

\vspace{0.25cm}
\centerline{\begin{minipage}{\textwidth}
        \resizebox{\textwidth}{!}{
		\begin{tcolorbox}
			{\small\texttt{(``large language model'' OR LLM* OR ``prompt engineering'' OR ``prompt tuning'' OR prompt*) AND (``requirements elicitation'' OR ``requirements analysis'' OR ``requirements specification'' OR ``requirements modeling'' OR ``requirements validation'' OR ``requirements verification'' OR ``requirements management'' OR ``requirements traceability'' OR ``requirements classification'' OR ``requirements document'' OR ``requirements engineering'')}}
		\end{tcolorbox}
        }
\end{minipage}}
\vspace{0.25cm}

We used LLM-representative terms in conjunction (\texttt{AND}) with RE-related terms. In the first set, we also considered the terms associated with prompt engineering, as this is a primary element that distinguishes the use of LLMs from non-generative models, at least in the terminology typically adopted in RE. Furthermore, in preliminary searches, we identified several papers employing the term `prompt' in its different variations. For RE-related terms, we adopted the same ones used by Zhao et al.~\cite{zhao2021natural} in their SMS on NLP for RE. 

We executed the query on September 11, 2024, which returned 244 studies. We then screened abstract and titles for relevance. This resulted in 136 papers, which were then analysed according to the inclusion and exclusion criteria (see Table~\ref{tab:inclusion_exclusion}). After applying the criteria, 62 studies were kept for further analysis.

\textbf{Secondary Venue-Based Search.} To ensure broader coverage and minimise the risk of omitting relevant studies, we conducted a secondary search that specifically targeted key journals and conferences in RE. This approach helps address potential limitations in terminology, indexing, or database coverage. The search focused on primary studies published between 2020 to 2024 in major SE/RE conferences and journals. Table~\ref{tab:venues-search} illustartes the selected venues for manual search. On December 18, 2024, we perfomed the secondary search using publications from these venues. After removing duplicates, screening them, and after using the same inclusion and exclusion criteria as in the initial search, 12 additional papers were selected---the limited number confirms the quality of the search string adopted for primary search. Our literature search resulted in 74 studies that we read, analysed and categorised.

\begin{table}[H]
	\centering
    \scriptsize
	\caption{Selected conference proceedings and journals for manual search}
	\label{tab:venues-search}
     \resizebox{1\textwidth}{!}{
	\begin{tabular}{|l|l|}
		\hline
		\textbf{Venue} & \textbf{Abbr.} \\
		\hline
		Requirements Engineering Journal & REJ \\
		\hline
		IEEE Transactions on Software Engineering & TSE \\
		\hline
		ACM Transactions on Software Engineering and Methodology & 
		TOSEM \\
		\hline
		Information and Software Technology & IST \\
		\hline
		Journal of Systems and Software & JSS \\
		\hline
		Empirical Software Engineering & EMSE \\
		\hline
		International Requirements Engineering Conference & RE \\
		\hline
		Requirements Engineering: Foundation for Software Quality & REFSQ \\
		\hline
	\end{tabular}
    }
\end{table}

\begin{table}[H]
    \centering
    \scriptsize
    \caption{Inclusion and exclusion criteria for study selection}
    \label{tab:inclusion_exclusion}
    \resizebox{1\textwidth}{!}{
    \begin{tabular}{|p{2cm}|p{10cm}|}
        \hline
        \textbf{ID} & \textbf{Criteria} \\
        \hline
        \multicolumn{2}{|c|}{\textbf{Inclusion Criteria}} \\
        \hline
        I1 & Primary studies using LLMs to support RE tasks. \\
        \hline
        I2 & Peer-reviewed studies published at conferences, workshops or journals \\
        \hline
        I3 & Papers in \textit{CORE A*/A/B} conferences\footnotemark[1], \textit{Q1/Q2} Journals\footnotemark[2], and workshops in {CORE A*/A/B} conferences\footnotemark[3]. \\
        \hline
        \multicolumn{2}{|c|}{\textbf{Exclusion Criteria}} \\
        \hline
        E1 & Papers not written in English. \\
        \hline
        E2 & Papers using LLMs without the purpose of supporting RE. \\
        \hline
        E3 & Secondary and tertiary studies, short papers (less than 5pp) and book chapters. \\
        \hline
        E4 & Conceptual papers without empirical evaluations. \\
        \hline
    \end{tabular}
    }
\end{table}
\footnotetext[1]{\url{https://portal.core.edu.au/conf-ranks/}}

\footnotetext[2]{\url{https://www.scimagojr.com/journalrank.php}}

\footnotetext[3]{For B venues, we screened the workshop studies, and considered only those that had a quality comparable to those of workshops in A*/A venues.}




\subsection{Data Extraction and Synthesis}

The first and last authors created a data extraction form to collect detailed information from each of the selected studies, and created classification schemes for each data field in the form. The fields in the form mirrored the RQs, and can be mapped one to one with the elements of the conceptual representation presented in Sect.~\ref{sec:conceptualmodel}. 

To construct our classification schemes for each of the field, we adopted a structured yet adaptive approach to categorise various dimensions of our study. The scheme development process was informed by existing literature and enriched with our domain expertise to ensure both their alignment with the current state of research and their continued relevance. Specifically, if the literature included a classification scheme that was deemed appropriate for the specific data field, considering the retrieved literature, such classification scheme was adopted or minimally adapted considering possible variants emerging during the analysis of the papers.  Otherwise, if no representative scheme could be retrieved from the literature, a novel scheme was defined by the authors inductively. During paper analysis, narrative information was initially reported in the data extraction form to characterise the data field of interest. Then, this narrative information was thematically analysed and aggregated into classes by the first and last author. These classes formed the final data extraction scheme for the specific field. 


In Table~\ref{tab:extracted_fields}, we list the extracted fields, together with the associated RQs, and the reference to their classification schemes. These latter are reported directly in the results section, to make it easier for the reader to interpret the results. The last column of the table refers to the source of the classification scheme, if any. It reports 'ind.' when the inductive approach was adopted, or '-', in case no classification scheme was required due to the nature of the field.

\begin{table}[H]
    \centering
    \scriptsize
    \caption{Data extraction form. Sch. = Classification scheme or Table with results, in case the scheme was not required. Ref. = Reference for the scheme definition. 'ind' = defined by induction. '-'= not required or trivial.}
    \label{tab:extracted_fields}
    \resizebox{1\textwidth}{!}{
    \begin{tabular}{|p{2.8cm}|p{7.8cm}|p{1cm}|p{1cm}|}
        \hline
        \textbf{Field} & \textbf{Description} & \textbf{Sch.} & \textbf{Ref.} \\
        \hline
        \multicolumn{4}{|c|}{\textbf{RQ1: Demographics}} \\ \hline
        Authors, Title, Year, Citation key, Venue Type, DOI & capture the bibliographic information of each study, including its identity, publication details, and contributing researchers. In our mapping study we use only year and venue type, i.e., conference/journal/workshop for classification purposes. & - & - \\ \hline
        \multicolumn{4}{|c|}{\textbf{RQ2: RE Usage}} \\ \hline
        RE Task & defines the RE task being addressed (e.g., elicitation, validation, traceability). & \ref{tab:combined_re_tasks_nlp} & \cite{ferrariappendix} \\ \hline
        RE Phase & refers to the RE lifecycle stage covered (e.g., elicitation, specification, analysis, management). & \ref{tab:combined_re_phase} & \cite{zhao2021natural} \\ \hline
        Input Artifact & identifies the data or artifacts used as input (e.g., requirements specifications, user stories, models) for the LLM. & \ref{tab:combined_input_artifacts} & \cite{zhao2021natural} \\ \hline
        Output Artifact & specifies the type of output generated (e.g., structured requirements, traceability links, classification labels). & \ref{tab:combined_output_artifacts} & \cite{zhao2021natural} \\ \hline
        \multicolumn{4}{|c|}{\textbf{RQ3: Prompt Engineering}} \\ \hline
        Prompting Technique & describes the general strategies used for prompting (e.g., zero-shot, few-shot, chain-of-thought). & \ref{tab:combined_prompting_strategies} & ind. \\ \hline
        Prompt Selection Strategy & explains how initial prompts are chosen (e.g., predefined templates, trial and error). & \ref{tab:combined_prompt_selection} & ind. \\ \hline
        Prompt Engineering Guidelines & outlines any frameworks, best practices, or structured guidance applied in prompt engineering. & \ref{tab:guideline_mapping} & ind. \\ \hline
        \multicolumn{4}{|c|}{\textbf{RQ4: Resources}} \\ \hline
        Types of LLM & lists the specific LLMs utilized in the study (e.g., GPT, Llama). & \ref{tab:llm_family_summary} & ind. \\ \hline
        Availability Link & link to the source code and/or prompts  shared for reproducibility purposes. & \ref{tab:code_availability} & - \\ \hline
        Dataset & describes the dataset(s) used, including whether they are public or proprietary and how they are referenced. & \ref{tab:datasets} & - \\ \hline
        Tool/Solution & specifies the specific name or identifier of the solution/tool, if available. & \ref{tab:tools_combined} & - \\ \hline
        \multicolumn{4}{|c|}{\textbf{RQ5: Evaluation}} \\ \hline
        Empirical Strategy & indicates what type of empirical strategy was adopted in the study, (e.g., laboratory experiment, experimental simulation, etc). & \ref{tab:evaluation_methods_combined} & \cite{stol2018abc} \\ \hline
        Output Assessment & describes how the outputs generated by the LLM are evaluated or interpreted, (e.g., qualitative methods, quantitative methods, or both). & \ref{tab:llm_evaluation_combined} & ind. \\ \hline
    \end{tabular}
    }
\end{table}

Data extraction was performed by the first author. Similarly to the search phase, we did not systematically check the reliability of the data extraction, but the first author consulted with the last author in case of doubts. A more systematic cross-check will be performed in future versions of the paper when this research will be submitted to a journal.

\section{Results}



We now present the results of our mapping study, addressing research questions RQ1–RQ5. These questions were answered using information collected from primary studies and by adopting classification schemas (see Sect.~\ref{sec:methodology}).

\subsection{RQ1: Demographics}
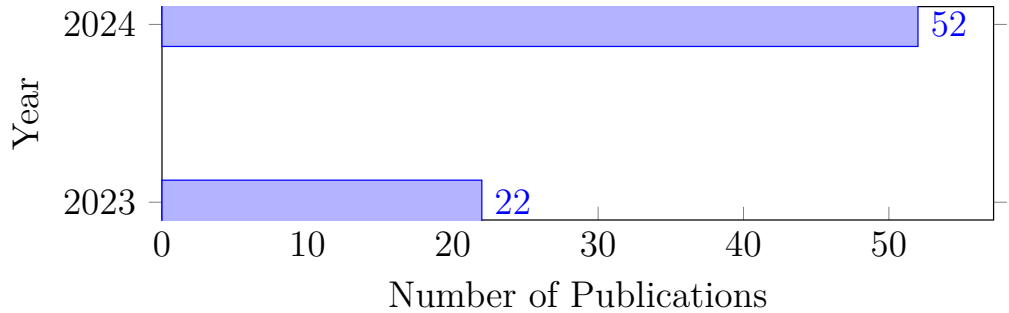
\begin{figure}[h]
	\centering
    \resizebox{\textwidth}{!}{
	\begin{tikzpicture}
		\begin{axis}[
			width=11cm,
			height=4cm,
			xbar,
			ylabel={Year},
			xlabel={Number of Publications},
			symbolic y coords={2023, 2024},
			ytick=data,
			xmin=0,
			bar width=0.5cm,
			nodes near coords,
			nodes near coords align={horizontal}
			]
			\addplot coordinates {(22,2023) (52,2024)};
		\end{axis}
	\end{tikzpicture}
    }
	\caption{Number of publications per year}
	\label{fig:publications}
\end{figure}

A total of 74 primary studies on the use of LLMs for requirements engineering were published between 2023 and 2024. Although the search time had no boundary, no paper in the field referred to LLMs before 2023 according to our search results. Despite the fact that generative pre-trained transformers already existed already in 2018 with GPT-1, we argue that RE researchers started getting interested in LLMs when the ChatGPT phenomenon exploded in 2023---it was released on November 30, 2022. This suggests that the RE community, much like the general public, became enthusiastic about LLMs primarily when ChatGPT drew widespread attention to the technology. Unlike earlier models, which demanded significant technical expertise to access and operate, ChatGPT provided an easy-to-use interface that suddenly made advanced generative capabilities available to anyone, including RE researchers. Fig.~\ref{fig:publications} shows the annual distribution of these publications. The number of studies increased significantly in 2024 compared to 2023 by 30 studies (a 136\% rise), indicating growing research interest in the LLM4RE area. Interestingly, the total number of papers published in the entire history of NLP for RE (NLP4RE), dating back to 1983 and surveyed by Zhao until 2019, is 404. In just two years, publications in LLM4RE already account for almost a fifth of that number. If this trend continues, by 2026, the number of LLM4RE papers could equal the total NLP4RE publications accumulated over roughly 40 years.

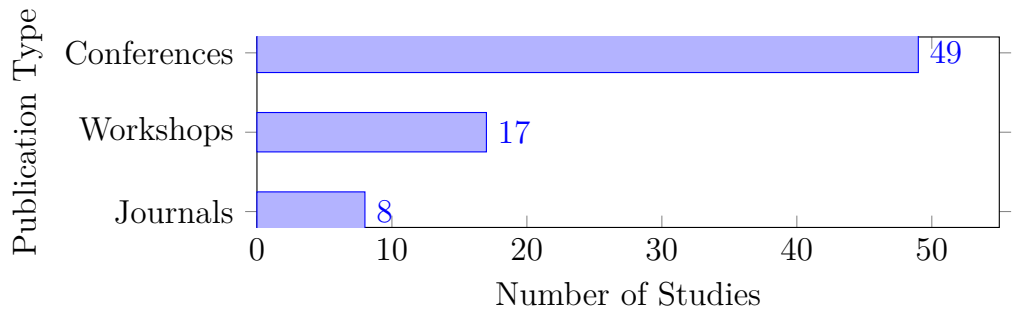
\begin{figure}[h]
	\centering
    \resizebox{\textwidth}{!}{
	\begin{tikzpicture}
		\begin{axis}[
			xbar, 
			width=11cm, 
			height=4cm, 
			ylabel={Publication Type},
			xlabel={Number of Studies},
			symbolic y coords={Journals, Workshops, Conferences}, 
			ytick=data,
			xmin=0,
			xmax=55, 
			bar width=0.5cm,
			nodes near coords
			]
			\addplot coordinates {(49,Conferences) (17,Workshops) (8,Journals)};
		\end{axis}
	\end{tikzpicture}
    }
	\caption{Distribution of research papers per venue type.}
	\label{fig:publication_distribution}
\end{figure}

The primary studies are distributed across three types of publication venues: conferences, workshops, and journals. Fig.~\ref{fig:publication_distribution} presents the numerical distribution of these studies by venue type. Most publications appeared in conferences (49 studies, 66\%), followed by workshops (17 studies, 23\%) and journals (8 studies, 11\%), indicating that LLM4RE is still in an exploratory stage, with findings more often presented in conferences and workshops than in archival journals.

Table~\ref{tab:venues} presents the top venues where more than two papers appeared. We see that the majority of the papers are at the flagship RE conference, i.e., RE (21 studies, 28\%), followed by the conference's workshop (15 studies, 20\%). However, a large part of the studies (32 studies, 43\%) appear in one or two venues only---not shown in the table. These venues include also HCII (International Conference on Human–Computer Interaction), LREC (Language Resources and Evaluation Conference), COLING (International Conference on Computational Linguistics), MSR (Mining Software Repositories), and CAV (International Conference on Computer-Aided Verification), which are not typical RE venues, and some even not SE ones. This indicates that there is a growing interest in requirements-related tasks and artifacts, even outside traditional RE and SE circles. It represents an opportunity for RE researchers to reach out to a broader set of communities--—such as AI, NLP, and HCI—--both to showcase domain-specific challenges and to benefit from cross-fertilization with methods and perspectives developed outside the RE field.

\begin{table}[H]
    \centering
    \scriptsize
    \caption{Most common venues (\# = number of studies).}
    \label{tab:venues}
     \resizebox{1\textwidth}{!}{%
    \begin{tabular}{|p{12cm}|p{1cm}|}
        \hline
        \textbf{Venue} & \textbf{\# } \\
        \hline
        IEEE International Requirements Engineering Conference (RE) & 21 \\
        \hline
        Requirements Engineering Conference Workshops (REW) & 15 \\
        \hline
        IEEE Transactions on Software Engineering (TSE) & 3 \\
        \hline
        Model-Driven Requirements Engineering (MoDRE) & 3 \\
        \hline
        Other Venue & 32 \\
        \hline
    \end{tabular}
    }
\end{table}

\begin{tcolorbox}[title=RQ1: Demographics]
	\begin{itemize}
    \item The surge of 74 studies starting only in 2023 shows that the RE interest in LLMs began only after the ChatGPT release. RE researchers, like the general public, engaged with LLMs once easy-to-use interfaces lowered technical barriers.  
    \item Publications rose 136\% in 2024 vs 2023, signaling rapidly growing momentum. In just two years, LLM4RE papers reached $\sim$20\% of all historical NLP4RE work (1983--2019). If growth continues, LLM4RE output may equal NLP4RE by 2026.  
    \item Most studies appear in conferences (66\%), fewer in workshops (23\%) and journals (11\%), indicating LLM4RE as an early-stage research area, with findings disseminated quickly through conferences rather than journals.   
    \item The flagship RE conference (28\%) dominates, but 57\% of papers are scattered across diverse venues. Papers also appear in non-RE/SE venues (e.g., HCII, LREC-COLING, MSR, CAV), showing cross-disciplinary interest. This trend is an opportunity for RE researchers to engage with broader communities (AI, NLP, HCI) for cross-fertilization of ideas.  
\end{itemize}
\end{tcolorbox}

\subsection{RQ2: RE Usage}
In this section, we explore how LLMs have been used in RE, considering different tasks, phases, as well input and output artifacts processed.

\subsubsection{RQ2.1: RE  Tasks}
Table~\ref{tab:combined_re_tasks_nlp} presents a list of tasks and their descriptions in our study, along with a mapping to the 74 studies that investigated them, including the number and percentage of studies for each task. The tasks are categorized into eleven categories, with each row listing the relevant references, the number of studies, and the percentage of the total they represent. The most frequently studied tasks are \textit{Requirements Elicitation} and \textit{Requirements Validation} (15 studies, 20\% each), indicating strong research attention toward both the initial gathering and the later evaluation of requirements. Other well-represented areas include \textit{Software Engineering Tasks} (11 studies, 15\%), \textit{Requirements Modelling} (9 studies, 12\%), and \textit{Requirements Classification} (8 studies, 11\%), highlighting the importance of both the structural and analytical aspects of RE. Conversely, tasks such as \textit{Requirements Retrieval} and \textit{Terminology Extraction} are each addressed by only one study (1\%), pointing to underexplored areas with opportunities for further investigation. 

While a direct comparison with NLP4RE studies is not straightforward---since Zhao et al.~\cite{zhao2021natural} employed a different classification scheme---some interesting patterns emerge. In their survey, the most common tasks were Detection (i.e., \textit{Requirements Defect Detection} in our categorization) and Classification. In contrast, in the LLM4RE landscape, these tasks rank only 8th and 5th, respectively. Instead, \textit{Elicitation}, \textit{Validation}, and \textit{Modelling}—--arguably the most human-intensive and inherently challenging tasks in RE—--have become the focal point, reflecting the unique capabilities of LLMs to support more cognitively demanding activities. This shift suggests that LLMs are not merely automating traditional NLP4RE tasks but are enabling research to tackle tasks that were previously considered difficult for machines, potentially reshaping the priorities and opportunities in RE research. In this context, it is also noteworthy that a substantial number of studies focus on broader \textit{Software Engineering Tasks}—--for example, using requirements to generate test cases. This trend indicates that LLMs provide the RE field with the opportunity to move beyond its traditional boundaries and explore intersections with other software engineering sub-disciplines.


\begin{table}[H]
	\centering
	\scriptsize
	\renewcommand{\arraystretch}{1.5}
	\setlength{\tabcolsep}{4pt}
	\caption{RE-Oriented NLP tasks: descriptions, references, and distribution}
	\label{tab:combined_re_tasks_nlp}
	\begin{tabular}{|p{4cm}|p{7.2cm}|c|c|}
		\hline
		\textbf{Task} & \textbf{Description / References} & \textbf{No.} & \textbf{\%} \\
		\hline
		
		\multirow{2}{=}{Requirements Elicitation}
		& \textbf{Description:} Gathering requirements from stakeholders through interviews, surveys, etc., to understand their needs, goals, and constraints. & \multirow{2}{*}{15} & \multirow{2}{*}{20\%} \\
		\cline{2-2}
		& \textbf{Studies:} \cite{Gorer2024-mu, Devathasan2024-tc, Fantechi2024-tt, Kolthoff2024-dt, Preda2024-xv, Nouri2024-jr, Chang2023-zg, Treude2023-er, Ronanki2023-sl, Jain2023-aq, Nikolaidis2024-ti, Mu2024-jl, Chen2023-pf, Marczak-Czajka2023-en, Xie2023-wv} & & \\
		\hline
		
		\multirow{2}{=}{Requirements Documentation}
		& \textbf{Description:} Recording requirements in structured formats such as SRS or user stories. & \multirow{2}{*}{4} & \multirow{2}{*}{5\%} \\
		\cline{2-2}
		& \textbf{Studies:} \cite{Lian2024-tk, Yang2024-ub, Li2024-dl, Ruan2023-pi} & & \\
		\hline
		
		\multirow{2}{=}{Requirements Defect Detection}
		& \textbf{Description:} Identifying inconsistencies or quality issues in requirements early in the lifecycle. & \multirow{2}{*}{4} & \multirow{2}{*}{4\%} \\
		\cline{2-2}
		& \textbf{Studies:} \cite{Fazelnia2024-hy, Lutze2024-ao, Silva2024-ap, Fantechi2023-gd} & & \\
		\hline
		
		\multirow{2}{=}{Requirements Prioritization}
		& \textbf{Description:} Ranking requirements by importance and feasibility to guide implementation order. & \multicolumn{2}{c|}{–} \\
		\hline
		
		\multirow{2}{=}{Requirements Tracing}
		& \textbf{Description:} Linking requirements across artifacts for traceability and impact analysis. & \multirow{2}{*}{5} & \multirow{2}{*}{7\%} \\
		\cline{2-2}
		& \textbf{Studies:} \cite{Abualhaija2024-vj, Feng2024-kj, Rejithkumar2024-zt, Gartner2024-hw, Rodriguez2023-wt} & & \\
		\hline
		
		\multirow{2}{=}{Requirements Classification}
		& \textbf{Description:} Organizing requirements (e.g., functional vs. non-functional) to aid analysis. & \multirow{2}{*}{8} & \multirow{2}{*}{11\%} \\
		\cline{2-2}
		& \textbf{Studies:} \cite{Fazelnia2024-hy, Feng2024-kj, Binder2024-gr, El-Hajjami2024-nc, Norheim2024-xn, Zhang2023-ml, Wei2023-ig, Marczak-Czajka2023-ew} & & \\
		\hline
		
		\multirow{2}{=}{Requirements Retrieval}
		& \textbf{Description:} Reusing relevant requirements from prior projects or documents. & \multirow{2}{*}{1} & \multirow{2}{*}{1\%} \\
		\cline{2-2}
		& \textbf{Studies:} \cite{Wei2024-nc} & & \\
		\hline
		
		\multirow{2}{=}{Requirements Terminology Extraction}
		& \textbf{Description:} Extracting domain-specific terms to create glossaries. & \multirow{2}{*}{1} & \multirow{2}{*}{1\%} \\
		\cline{2-2}
		& \textbf{Studies:} \cite{Krishna2024-ck} & & \\
		\hline
		
		\multirow{2}{=}{Requirements Relations Extraction}
		& \textbf{Description:} Identifying dependencies and associations between requirements. & \multicolumn{2}{c|}{–} \\
		\hline
		
		\multirow{2}{=}{Requirements Modeling}
		& \textbf{Description:} Representing requirements in abstract forms (e.g., UML diagrams). & \multirow{2}{*}{9} & \multirow{2}{*}{12\%} \\
		\cline{2-2}
		& \textbf{Studies:} \cite{Klievtsova2024-ep, Ferrari2024-pu, Alharbi2024-us, Bragilovski2024-cv, Rejithkumar2024-sv, Azeem2024-az, Cosler2023-uq, Zhang2023-xc, Arulmohan2023-xf} & & \\
		\hline
		
		\multirow{2}{=}{Requirements Validation}
		& \textbf{Description:} Ensuring that requirements reflect stakeholder goals and expectations. & \multirow{2}{*}{15} & \multirow{2}{*}{20\%} \\
		\cline{2-2}
		& \textbf{Studies:} \cite{Lubos2024-pw, Shahandashti2024-oo, Hey2024-wh, Hassine2024-by, Chen2024-mj, Sivakumar2024-jb, Uygun2024-ug, Karolita2024-rp, Singhal2024-wz, Veizaga2024-im, Cotroneo2024-yt, Bertram2023-av, Soares2023-xj, Gorgen2024-ql, Arora2016-nn} & & \\
		\hline
		
		\multirow{2}{=}{Change Impact Analysis}
		& \textbf{Description:} Assessing the effect of proposed changes on existing requirements. & \multicolumn{2}{c|}{–} \\
		\hline
		
		\multirow{2}{=}{Software Engineering Tasks}
		& \textbf{Description:} Applying requirements to support downstream SE tasks (e.g., testing, coding). & \multirow{2}{*}{11} & \multirow{2}{*}{15\%} \\
		\cline{2-2}
		& \textbf{Studies:} \cite{Arora2024-gs, Hasso2024-ge, Yaacov2024-xu, Khakzad_Shahandashti2024-bb, Li2024-zw, Ren2024-aj, Tikayat_Ray2023-wm, Yang2023-ru, Schafer2024-nf, Liao2024-my, North2024-td} & & \\
		\hline
		
		\multirow{2}{=}{Legal Requirements Analysis}
		& \textbf{Description:} Integrating legal/compliance constraints into RE through systematic analysis. & \multirow{2}{*}{4} & \multirow{2}{*}{4\%} \\
		\cline{2-2}
		& \textbf{Studies:} \cite{Abualhaija2024-vj, Hassani2024-ej, De_Kinderen2024-np, Gorgen2024-ql} & & \\
		\hline
	\end{tabular}
\end{table}

\begin{figure}[H]
    \centering
    \caption{RE tasks and references}
    \label{fig:re_tasks_histogram}
    \resizebox{\textwidth}{!}{
    \begin{tikzpicture}
        \begin{axis}[
            xbar,
            width=10cm, height=7cm,
            xlabel={References (\%)},
            ylabel={RE Tasks},
            symbolic y coords={
                Impact Analysis, Relations, Prioritization, Terminology, Retrieval, Legal Analysis, Defect Detection, Documentation, Tracing, Classification, Modeling, SE Tasks, Validation, Elicitation
            },
            ytick=data,
            yticklabel style={font=\scriptsize},
            bar width=0.35cm,
            nodes near coords={\pgfmathprintnumber{\pgfplotspointmeta}\%},
            nodes near coords style={
                /pgf/number format/.cd,
                fixed,
                fixed zerofill,
                precision=0,
                /tikz/.cd,
                font=\scriptsize
            },
            tick label style={font=\scriptsize},
            label style={font=\scriptsize},
            title style={font=\scriptsize},
            xmin=0, xmax=25
        ]
        \addplot coordinates {
            (20,Elicitation) (20,Validation) (15,SE Tasks) (12,Modeling) (11,Classification)
            (7,Tracing) (4,Documentation) (5,Defect Detection) (4,Legal Analysis) (1,Retrieval)
            (1,Terminology) (0,Prioritization) (0,Relations) (0,Impact Analysis)
        };
        \end{axis}
    \end{tikzpicture}
    }
\end{figure}



\subsubsection{RQ2.2: RE Phases}
Table~\ref{tab:combined_re_phase} presents six major RE phases, their definition and the distribution of primary studies across these phases, along with references to the studies. The studies are grouped into six high-level RE phases based on their primary focus. The most represented phases are \textit{Elicitation} and \textit{Validation \& Verification (V\&V)} (16 studies, 22\% each), indicating a strong emphasis in the literature on both the initial capture and the subsequent evaluation of requirements. The \textit{Analysis} phase is also well represented (14 studies, 19\%), reflecting active research on interpreting, refining, and classifying requirements. Both \textit{Modeling} and \textit{Management} are addressed by 9 studies each (12\%), suggesting steady but comparatively lower attention to the structuring and ongoing handling of requirements. The \textit{Other} category (12 studies, 16\%) includes studies focused on broader or cross-cutting concerns that do not align specifically with a single RE phase. Overall, this landscape suggests that  LLMs are not limited to specific stages but are being leveraged throughout the RE process, supporting both phase-specific and cross-cutting activities. 

Again, comparing our results with those of Zhao et al.~\cite{zhao2021natural} offers additional insights. In their survey, \textit{Analysis} was by far the most commonly addressed phase (42.7\%), while \textit{V\&V} accounted for only 4\% of the studies, and \textit{Elicitation} represented 16.76\%. In contrast, the landscape in LLM4RE is completely different: as observed for the tasks, the first and last phases of the RE lifecycle have gained prominence. On the one hand, this shift opens opportunities for automating highly relevant and time-consuming RE activities. On the other hand, caution is warranted, as these phases are particularly delicate—they involve direct interaction with stakeholders, and \textit{V\&V}, in particular, can be critical when compliance with standards or regulatory requirements is at stake. Thus, while LLMs offer significant potential, careful consideration of their role and limitations in sensitive phases of the RE process remains essential.

\begin{table}[H]
	\centering
	\scriptsize
	\renewcommand{\arraystretch}{1.5}
	\setlength{\tabcolsep}{4pt}
	\caption{RE phases used in this research (adapted from \cite{10.1145/3444689}): definitions, references, and frequencies}
	\label{tab:combined_re_phase}
	\begin{tabular}{|p{4.3cm}|p{7.2cm}|c|c|}
		\hline
		\textbf{RE Phase} & \textbf{Definition / References} & \textbf{No.} & \textbf{\%} \\
		\hline
		
		\multirow{2}{=}{Elicitation}
		& \textbf{Definition:} Activities that help understand goals, objectives, and stakeholder needs for building a proposed software system. & \multirow{2}{*}{16} & \multirow{2}{*}{22\%} \\
		\cline{2-2}
		& \textbf{Studies:} \cite{Gorer2024-mu, Devathasan2024-tc, Fantechi2024-tt, Mu2024-jl, Wei2024-nc, Kolthoff2024-dt, Preda2024-xv, Nouri2024-jr, Nikolaidis2024-ti, Chang2023-zg, Treude2023-er, Jain2023-aq, Ronanki2023-sl, Chen2023-pf, Xie2023-wv, Marczak-Czajka2023-en} & & \\
		\hline
		
		\multirow{2}{=}{Analysis}
		& \textbf{Definition:} Evaluating the quality of recorded requirements and identifying anomalies like ambiguity, inconsistency, or incompleteness. & \multirow{2}{*}{14} & \multirow{2}{*}{19\%} \\
		\cline{2-2}
		& \textbf{Studies:} \cite{Abualhaija2024-vj, Fazelnia2024-hy, Hassani2024-ej, Krishna2024-ck, Lutze2024-ao, Norheim2024-xn, De_Kinderen2024-np, Binder2024-gr, El-Hajjami2024-nc, Marczak-Czajka2023-ew, Wei2023-ig, Silva2024-ap, Fantechi2023-gd, Zhang2023-ml} & & \\
		\hline
		
		\multirow{2}{=}{Modeling}
		& \textbf{Definition:} Building conceptual models of requirements that are interpretable and structured. & \multirow{2}{*}{9} & \multirow{2}{*}{12\%} \\
		\cline{2-2}
		& \textbf{Studies:} \cite{Klievtsova2024-ep, Ferrari2024-pu, Alharbi2024-us, Bragilovski2024-cv, Rejithkumar2024-sv, Azeem2024-az, Zhang2023-xc, Cosler2023-uq, Arulmohan2023-xf} & & \\
		\hline
		
		\multirow{2}{=}{Management}
		& \textbf{Definition:} Managing requirements across time and versions, and ensuring traceability among requirements and downstream artifacts. & \multirow{2}{*}{9} & \multirow{2}{*}{12\%} \\
		\cline{2-2}
		& \textbf{Studies:} \cite{Abualhaija2024-vj, Lian2024-tk, Feng2024-kj, Rejithkumar2024-zt, Gartner2024-hw, Li2024-dl, Yang2024-ub, Ruan2023-pi, Rodriguez2023-wt} & & \\
		\hline
		
		\multirow{2}{=}{Validation \& Verification}
		& \textbf{Definition:} Ensuring requirements reflect stakeholder intent (validation) and proving that specifications meet them (verification), e.g., via model checking. & \multirow{2}{*}{16} & \multirow{2}{*}{22\%} \\
		\cline{2-2}
		& \textbf{Studies:} \cite{Lubos2024-pw, Shahandashti2024-oo, Chen2024-mj, Hey2024-wh, Hassine2024-by, Sivakumar2024-jb, Uygun2024-ug, Karolita2024-rp, Singhal2024-wz, Arora2016-nn, De_Kinderen2024-np, Gorgen2024-ql, Veizaga2024-im, Cotroneo2024-yt, Bertram2023-av, Soares2023-xj} & & \\
		\hline
		
		\multirow{2}{=}{Other}
		& \textbf{Definition:} Covers RE-related activities like test-case generation (``Testing’’) or deriving design artifacts (``Design’’). & \multirow{2}{*}{12} & \multirow{2}{*}{16\%} \\
		\cline{2-2}
		& \textbf{Studies:} \cite{Arora2024-gs, Hasso2024-ge, Yaacov2024-xu, Arora2024-ib, Khakzad_Shahandashti2024-bb, Li2024-zw, North2024-td, Ren2024-aj, Tikayat_Ray2023-wm, Yang2023-ru, Liao2024-my, Schafer2024-nf} & & \\
		\hline
	\end{tabular}
\end{table}

\begin{figure}[H]
    \centering
    \caption{RE phases and references}
    \label{fig:re_phases_histogram}
    \resizebox{\textwidth}{!}{
    \begin{tikzpicture}
        \begin{axis}[
            xbar,
            width=10cm, height=5cm,
            xlabel={References (\%)},
            ylabel={RE Phases},
            symbolic y coords={Modeling,Management,Other,Analysis,Elicitation,Validation \& Verification},
            ytick=data,
            xmin=0, xmax=25,
            bar width=0.25cm,
            nodes near coords={\pgfmathprintnumber{\pgfplotspointmeta}\%},
            tick label style={font=\scriptsize},
            label style={font=\scriptsize},
            legend style={font=\scriptsize},
            nodes near coords style={/pgf/number format/fixed,/pgf/number format/precision=0,font=\scriptsize}
        ]
        \addplot coordinates {
            (12,Modeling) (12,Management) (16,Other) (19,Analysis) (22,Elicitation) (22,Validation \& Verification)
        };
        \end{axis}
    \end{tikzpicture}
    }
\end{figure}

\subsubsection{RQ2.3: Input Artifacts}

Table~\ref{tab:combined_input_artifacts} presents the nine categories of input artifacts identified in the surveyed literature, along with the distribution of input artifacts used in the analyzed studies and references to the corresponding papers. The studies were categorized into nine artifact types based on the primary source of input they addressed. The most common artifact type is \textit{System/Software Requirements Specifications} (28 studies, 38\%), indicating that a majority of research in LLM4RE focuses on traditional, structured requirement documents. \textit{Issues and User Feedback} also account for a significant portion (9 studies, 12\%), highlighting interest in mining user-reported content from platforms such as issue trackers and product reviews. Other moderately represented types include \textit{Regulatory and Legal Artifacts} (7 studies, 9\%), \textit{Technical Documents and Manuals}, and \textit{Models and Ontologies} (6 studies, 8\% each), as well as \textit{Others} (8 studies, 11\%), the latter encompassing inputs that do not fit neatly into a single category. Less frequently addressed artifact types include \textit{Specific Quality Requirements} and \textit{User Stories and Agile Requirements} (4 studies, 5\% each), and \textit{Elicitation Documents} (2 studies, 3\%). This distribution highlights the dominance of formal requirements specifications in current research, while also pointing to a growing interest in utilizing informal and diverse artifact sources for RE tasks. Also in Zhao et al.~\cite{zhao2021natural},  the majority of the studies (60\%) used \textit{System/Software Requirements Specifications} as input artifacts. Here, the percentage has substantially decreased and there is a slightly more balanced distribution, suggesting that LLMs enable tackling not only novel tasks and phases but also interpreting novel types of documents.


\begin{table}[H]
	\centering
	\scriptsize
	\renewcommand{\arraystretch}{1.5}
	\setlength{\tabcolsep}{4pt}
	\caption{Input artifact categories: definitions, references, and frequencies}
	\label{tab:combined_input_artifacts}
    \resizebox{1\textwidth}{!}{
	\begin{tabular}{|p{4.5cm}|p{7.2cm}|c|c|}
		\hline
		\textbf{Category} & \textbf{Definition / References} & \textbf{No.} & \textbf{\%} \\
		\hline
		
		\multirow{2}{=}{Regulatory and Legal Artifacts}
		& \textbf{Definition:} Documents defining compliance, policies, contracts, and legal obligations for systems or projects. & \multirow{2}{*}{7} & \multirow{2}{*}{9\%} \\
		\cline{2-2}
		& \textbf{Studies:} \cite{Abualhaija2024-vj, Hassani2024-ej, Rejithkumar2024-zt, Singhal2024-wz, Azeem2024-az, Ren2024-aj, Jain2023-aq} & & \\
		\hline
		
		\multirow{2}{=}{System/Software Requirements Specifications}
		& \textbf{Definition:} Formalized documents listing the functional and non-functional requirements of a system. & \multirow{2}{*}{28} & \multirow{2}{*}{38\%} \\
		\cline{2-2}
		& \textbf{Studies:} \cite{Fazelnia2024-hy, Lian2024-tk, Arora2024-gs, Lubos2024-pw, Hasso2024-ge, Shahandashti2024-oo, Ferrari2024-pu, Yaacov2024-xu, Fantechi2024-tt, Hey2024-wh, Krishna2024-ck, Mu2024-jl, Gartner2024-hw, North2024-td, Wei2024-nc, Uygun2024-ug, Preda2024-xv, Nouri2024-jr, Rejithkumar2024-sv, Norheim2024-xn, Arora2016-nn, De_Kinderen2024-np, Chang2023-zg, Cosler2023-uq, Ruan2023-pi, Fantechi2023-gd, Bertram2023-av, Veizaga2024-im} & & \\
		\hline
		
		\multirow{2}{=}{Specific Quality Requirements}
		& \textbf{Definition:} Requirements focusing on a specific system quality dimension, e.g., security or safety. & \multirow{2}{*}{4} & \multirow{2}{*}{5\%} \\
		\cline{2-2}
		& \textbf{Studies:} \cite{Feng2024-kj, Khakzad_Shahandashti2024-bb, Sivakumar2024-jb, Marczak-Czajka2023-en} & & \\
		\hline
		
		\multirow{2}{=}{Elicitation Documents}
		& \textbf{Definition:} Elicitation artifacts such as interview transcripts, focus groups, and workshops. & \multirow{2}{*}{2} & \multirow{2}{*}{3\%} \\
		\cline{2-2}
		& \textbf{Studies:} \cite{Gorer2024-mu, Ronanki2023-sl} & & \\
		\hline
		
		\multirow{2}{=}{User Stories and Agile Requirements}
		& \textbf{Definition:} Lightweight, iterative requirement descriptions commonly used in Agile methodologies. & \multirow{2}{*}{4} & \multirow{2}{*}{5\%} \\
		\cline{2-2}
		& \textbf{Studies:} \cite{Li2024-zw, Bragilovski2024-cv, Kolthoff2024-dt, Arulmohan2023-xf} & & \\
		\hline
		
		\multirow{2}{=}{Technical Documents and Manuals}
		& \textbf{Definition:} Detailed technical documentation for implementation, usage, and maintenance. & \multirow{2}{*}{6} & \multirow{2}{*}{8\%} \\
		\cline{2-2}
		& \textbf{Studies:} \cite{Chen2024-mj, Arora2024-ib, Lutze2024-ao, Liao2024-my, Schafer2024-nf, Soares2023-xj} & & \\
		\hline
		
		\multirow{2}{=}{Issues and User Feedback}
		& \textbf{Definition:} App reviews, crowd requirements, and issues from platforms such as GitHub, Jira, etc. & \multirow{2}{*}{9} & \multirow{2}{*}{12\%} \\
		\cline{2-2}
		& \textbf{Studies:} \cite{Devathasan2024-tc, Karolita2024-rp, Zhang2023-xc, Gorgen2024-ql, El-Hajjami2024-nc, Yang2024-ub, Tikayat_Ray2023-wm, Wei2023-ig, Zhang2023-ml} & & \\
		\hline
		
		\multirow{2}{=}{Models and Ontologies}
		& \textbf{Definition:} Conceptual models and ontologies aiding requirements engineering and analysis. & \multirow{2}{*}{6} & \multirow{2}{*}{8\%} \\
		\cline{2-2}
		& \textbf{Studies:} \cite{Klievtsova2024-ep, Hassine2024-by, Alharbi2024-us, Nikolaidis2024-ti, Yang2023-ru, Xie2023-wv} & & \\
		\hline
		
		\multirow{2}{=}{Others}
		& \textbf{Definition:} Other relevant artifacts or processes not categorized above. & \multirow{2}{*}{8} & \multirow{2}{*}{11\%} \\
		\cline{2-2}
		& \textbf{Studies:} \cite{Li2024-dl, Binder2024-gr, Treude2023-er, Marczak-Czajka2023-ew, Chen2023-pf, Cotroneo2024-yt, Silva2024-ap, Rodriguez2023-wt} & & \\
		\hline
	\end{tabular}
    }
\end{table}

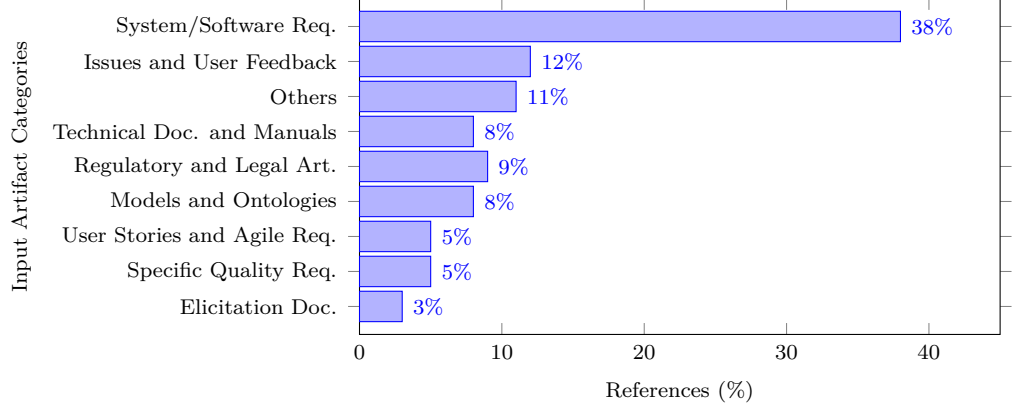
\begin{figure}[H]
    \centering
    \caption{Input artifact categories and references}
    \label{fig:combined_input_artifacts}
    \resizebox{\textwidth}{!}{
    \begin{tikzpicture}
        \begin{axis}[
            xbar,
            width=10cm, height=6cm,
            xlabel={References (\%)},
            ylabel={Input Artifact Categories},
            symbolic y coords={
                Elicitation Doc.,
                Specific Quality Req.,
                User Stories and Agile Req.,
                Models and Ontologies,
                Regulatory and Legal Art.,
                Technical Doc. and Manuals,
                Others,
                Issues and User Feedback,
                System/Software Req.
            },
            ytick=data,
            yticklabel style={font=\scriptsize},
            bar width=0.4cm,
            nodes near coords={\pgfmathprintnumber{\pgfplotspointmeta}\%},
            nodes near coords style={/pgf/number format/fixed,/pgf/number format/precision=0,font=\scriptsize},
            tick label style={font=\scriptsize},
            label style={font=\scriptsize},
            title style={font=\scriptsize},
            xmin=0, xmax=45
        ]
        \addplot coordinates {
            (3,Elicitation Doc.)
            (5,Specific Quality Req.)
            (5,User Stories and Agile Req.)
            (8,Models and Ontologies)
            (9,Regulatory and Legal Art.)
            (8,Technical Doc. and Manuals)
            (11,Others)
            (12,Issues and User Feedback)
            (38,System/Software Req.)
        };
        \end{axis}
    \end{tikzpicture}
    }
\end{figure}

\subsubsection{RQ2.4: Output Artifacts}

Table~\ref{tab:combined_output_artifacts} presents the categorization of output artifacts identified in the primary studies, divided into ten types based on their nature. It also shows the distribution of these artifacts across the surveyed studies, along with references to the corresponding papers. The most frequently targeted artifact type is \textit{System/Software Requirements Specifications} (15 studies, 20\%), similar to the case of input artifacts. \textit{Requirements Analysis Reports} are also highly represented (14 studies, 19\%), indicating a strong focus on analytic---i.e., not purely generative---activities such as classification, refinement, and prioritization of requirements. \textit{Models and Designs} appear in 10 studies (14\%), in line with the increasing interest in model generation tasks. \textit{Compliance Documents and Legal Artifacts} (8 studies, 11\%) and \textit{User Profiles} (7 studies, 9\%) highlight efforts to address domain-specific and user-centric concerns. Several artifact types are addressed in a moderate number of studies, including \textit{Code Artifacts} (6 studies, 8\%), \textit{Testing Artifacts} and \textit{Traceability Artifacts} (4 studies, 5\% each), and \textit{Interview Scripts and Questionnaires} (4 studies, 5\%). \textit{Process Descriptions} are the least represented (2 studies, 3\%). Overall, the table reveals that while much of the research focuses on core RE documents and analysis outputs, there is a growing effort to apply LLMs to a broader variety of RE outputs. In the study of Zhao et al.~\cite{zhao2021natural}, output artifacts were not present. This marks the main difference between NLP4RE studies and LLM4RE studies: with LLM4RE, we can tackle novel, generative tasks that were not even possible before. At the same time, we can still tackle analytic tasks (i.e., those associated with \textit{Requirements Analysis Reports}), showing the high versatility of LLM-based solutions. 

\begin{table}[H]
	\centering
	\scriptsize
	\renewcommand{\arraystretch}{1.5}
	\setlength{\tabcolsep}{4pt}
	\caption{Output artifact categories: definitions, references, and frequencies}
	\label{tab:combined_output_artifacts}
     \resizebox{1\textwidth}{!}{%
	\begin{tabular}{|p{3cm}|p{8.5cm}|c|c|}
		\hline
		\textbf{Category} & \textbf{Definition / References} & \textbf{No.} & \textbf{\%} \\
		\hline
		
		\multirow{2}{=}{System/Software Requirements Specifications} 
		& \textbf{Definition:} Formalized documents listing the functional and non-functional requirements of a system. & \multirow{2}{*}{15} & \multirow{2}{*}{20\%} \\
		\cline{2-2}
		& \textbf{Studies:} \cite{Lian2024-tk, Hasso2024-ge, Shahandashti2024-oo, Krishna2024-ck, Li2024-zw, Lutze2024-ao, Norheim2024-xn, Binder2024-gr, El-Hajjami2024-nc, Yang2024-ub, Tikayat_Ray2023-wm, Cosler2023-uq, Veizaga2024-im, Silva2024-ap, Bertram2023-av} & & \\
		\hline
		
		\multirow{2}{=}{Process Descriptions}
		& \textbf{Definition:} Textual descriptions of processes, methods, or procedures involved in system development. & \multirow{2}{*}{2} & \multirow{2}{*}{3\%} \\
		\cline{2-2}
		& \textbf{Studies:} \cite{Klievtsova2024-ep, Gorgen2024-ql} & & \\
		\hline
		
		\multirow{2}{=}{Models and Designs}
		& \textbf{Definition:} Visual or structural representations (e.g., diagrams, models) that support understanding or system design. & \multirow{2}{*}{10} & \multirow{2}{*}{14\%} \\
		\cline{2-2}
		& \textbf{Studies:} \cite{Ferrari2024-pu, Chen2024-mj, Bragilovski2024-cv, Arora2016-nn, De_Kinderen2024-np, Ren2024-aj, Ruan2023-pi, Arulmohan2023-xf, Chen2023-pf, Xie2023-wv} & & \\
		\hline
		
		\multirow{2}{=}{Code Artifacts}
		& \textbf{Definition:} Source code and deliverables implementing system functions or refined from earlier iterations. & \multirow{2}{*}{6} & \multirow{2}{*}{8\%} \\
		\cline{2-2}
		& \textbf{Studies:} \cite{Yaacov2024-xu, Mu2024-jl, Wei2024-nc, Nikolaidis2024-ti, Chang2023-zg, Liao2024-my} & & \\
		\hline
		
		\multirow{2}{=}{Requirements Analysis Reports}
		& \textbf{Definition:} Reports analyzing consistency, contradictions, coverage, or other aspects of requirements. & \multirow{2}{*}{14} & \multirow{2}{*}{19\%} \\
		\cline{2-2}
		& \textbf{Studies:} \cite{Fazelnia2024-hy, Lubos2024-pw, Fantechi2024-tt, Feng2024-kj, Khakzad_Shahandashti2024-bb, Gartner2024-hw, Uygun2024-ug, Preda2024-xv, Zhang2023-xc, Li2024-dl, Ronanki2023-sl, Cotroneo2024-yt, Fantechi2023-gd, Soares2023-xj} & & \\
		\hline
		
		\multirow{2}{=}{Testing Artifacts}
		& \textbf{Definition:} Test cases, mutation testing, prototypes, or validation-related documents. & \multirow{2}{*}{4} & \multirow{2}{*}{5\%} \\
		\cline{2-2}
		& \textbf{Studies:} \cite{Arora2024-gs, Kolthoff2024-dt, Yang2023-ru, Schafer2024-nf} & & \\
		\hline
		
		\multirow{2}{=}{Traceability Artifacts}
		& \textbf{Definition:} Documents linking requirements to code or other outputs to support impact and change analysis. & \multirow{2}{*}{4} & \multirow{2}{*}{5\%} \\
		\cline{2-2}
		& \textbf{Studies:} \cite{Hey2024-wh, Hassine2024-by, North2024-td, Rodriguez2023-wt} & & \\
		\hline
		
		\multirow{2}{=}{Compliance Documents and Legal Artifacts}
		& \textbf{Definition:} Legal/compliance documents such as GDPR checks, safety cases, contracts, or regulations. & \multirow{2}{*}{8} & \multirow{2}{*}{11\%} \\
		\cline{2-2}
		& \textbf{Studies:} \cite{Abualhaija2024-vj, Hassani2024-ej, Rejithkumar2024-zt, Arora2024-ib, Sivakumar2024-jb, Nouri2024-jr, Azeem2024-az, Jain2023-aq} & & \\
		\hline
		
		\multirow{2}{=}{Interview Scripts and Questionnaires}
		& \textbf{Definition:} Structured elicitation tools like interview guides, question flows, and surveys. & \multirow{2}{*}{4} & \multirow{2}{*}{5\%} \\
		\cline{2-2}
		& \textbf{Studies:} \cite{Gorer2024-mu, Alharbi2024-us, Rejithkumar2024-sv, Singhal2024-wz} & & \\
		\hline
		
		\multirow{2}{=}{User Profiles}
		& \textbf{Definition:} Representations of user needs or values (e.g., personas, empathy maps, value stories). & \multirow{2}{*}{7} & \multirow{2}{*}{9\%} \\
		\cline{2-2}
		& \textbf{Studies:} \cite{Devathasan2024-tc, Karolita2024-rp, Treude2023-er, Marczak-Czajka2023-ew, Wei2023-ig, Zhang2023-ml, Marczak-Czajka2023-en} & & \\
		\hline
		
	\end{tabular}
    }
\end{table}

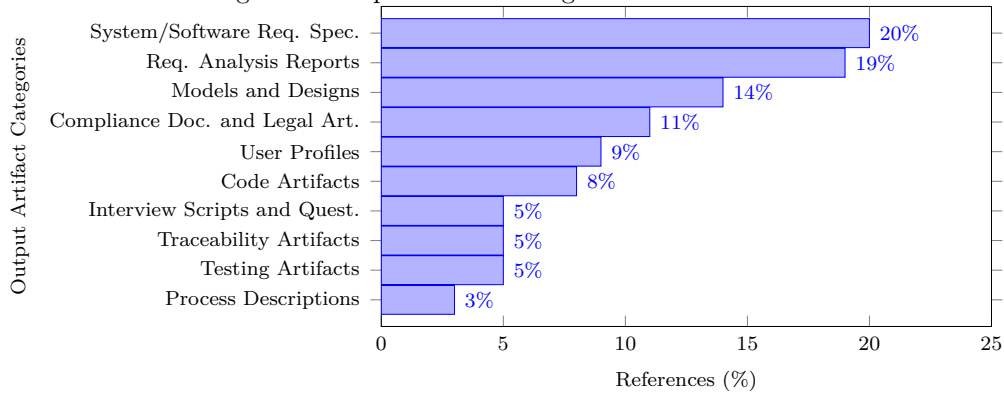
\begin{figure}[H]
    \centering
    \caption{Output artifact categories and references}
    \label{fig:combined_output_artifacts}
    \resizebox{\textwidth}{!}{
    \begin{tikzpicture}
        \begin{axis}[
            xbar,
            width=10cm, height=6cm,
            xlabel={References (\%)},
            ylabel={Output Artifact Categories},
            symbolic y coords={
                Process Descriptions,
                Testing Artifacts,
                Traceability Artifacts,
                Interview Scripts and Quest.,
                Code Artifacts,
                User Profiles,
               Compliance Doc. and Legal Art.,
                Models and Designs,
                Req. Analysis Reports,
                System/Software Req. Spec.
            },
            ytick=data,
            yticklabel style={font=\scriptsize},
            bar width=0.4cm,
            nodes near coords={\pgfmathprintnumber{\pgfplotspointmeta}\%},
            nodes near coords style={
                /pgf/number format/.cd,
                fixed,
                fixed zerofill,
                precision=0,
                /tikz/.cd,
                font=\scriptsize
            },
            tick label style={font=\scriptsize},
            label style={font=\scriptsize},
            title style={font=\scriptsize},
            xmin=0, xmax=25
        ]
        \addplot coordinates {
            (3,Process Descriptions)
            (5,Testing Artifacts)
            (5,Traceability Artifacts)
            (5,Interview Scripts and Quest.)
            (8,Code Artifacts)
            (9,User Profiles)
            (11,Compliance Doc. and Legal Art.)
            (14,Models and Designs)
            (19,Req. Analysis Reports)
        (20,System/Software Req. Spec.)
        };
        \end{axis}
    \end{tikzpicture}
    }
\end{figure}

\subsubsection{RQ2.5: Input and Output Artifacts Mapping}



\begin{figure}[H]
    \centering
    \includegraphics[width=1\textwidth]{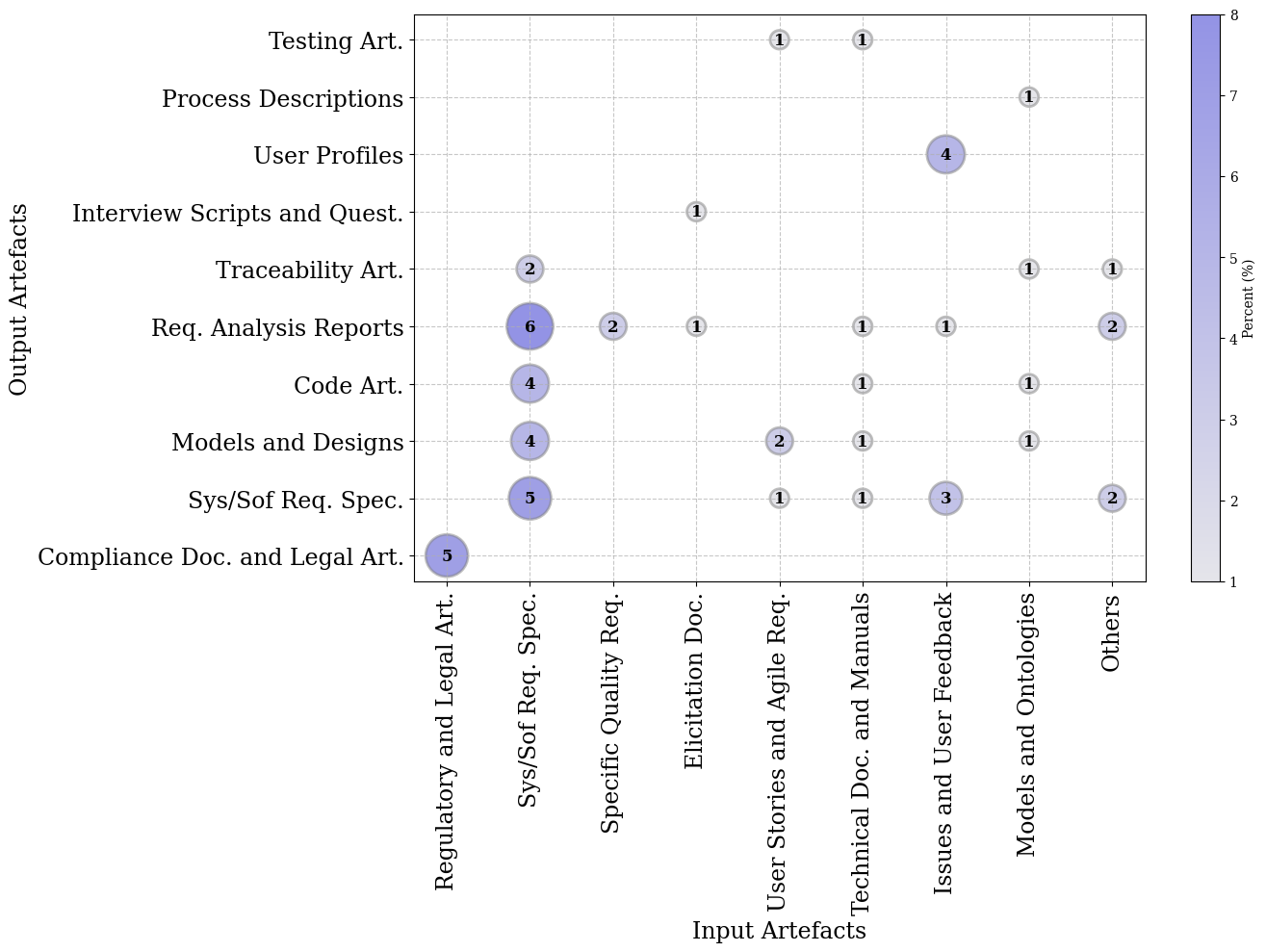}
    \caption{Bubble plot of input-output artifact mappings across primary studies.}
    \label{fig:input_output_bubble}
\end{figure}

We analyze the co-occurrence between input and output artifacts across the reviewed studies to identify common pairs. Table~\ref{tab:common_papers_grouped_refs} summarizes these pairs, while Fig.~\ref{fig:input_output_bubble} shows a bubble plot of the co-occurrences. 
The most evident aspect is the high number of pairs (27 across 74 studies), which points to a fragmented research landscape pursuing widely different goals. This reflects the limited maturity of the field and highlights its exploratory nature, where academics are primarily investigating novel application opportunities in RE rather than converging on a set of well-defined objectives. On the other hand, it also highlights the versatility of the field, which can be seen as a playground for experimenting with diverse applications.

The most frequent pair is \textit{System/Software Requirements Specifications}-\textit{Requirements Analysis Reports} (6 studies, 8\%). This confirms that the interest in purely analytic tasks is still present. The pair is followed by \textit{Regulatory and Legal Artifacts}-\textit{Compliance Documents and Legal Artifacts}. Regulatory compliance in RE is a lively research area, especially since software artifacts are being increasingly regulated, e.g., by GDPR\footnote{\url{https://eur-lex.europa.eu/legal-content/EN/TXT/?uri=celex:32016R0679}}, or used as part of mission-critical products, such as healthcare devices, and transport systems that need to abide by high safety standards. We argue that, with the novel regulations such as the AI Act\footnote{\url{https://digital-strategy.ec.europa.eu/en/policies/regulatory-framework-ai}}, these types of studies will become more and more popular. Interestingly, another common co-occurrence is \textit{System/Software Requirements Specifications}-\textit{System/Software Requirements Specifications} (5 studies, 7\%). These studies typically focus on enhancing specifications, for example,  by improving requirements completeness through question-answering or through automatic completion, or by translating requirements into well-defined templates. Despite the strong push toward exploring new applications, the fundamental concern for requirements quality continues to be at the core of RE research.



\begin{table}[H]
	\centering
	\scriptsize
	\renewcommand{\arraystretch}{1.2}
	\setlength{\tabcolsep}{4pt}
	\resizebox{\textwidth}{!}{%
		\begin{tabular}{|p{4cm}|p{5cm}|p{3cm}|c|c|}
			\hline
			\textbf{Input Artifact} & \textbf{Output Artifact} & \textbf{Papers} & \textbf{No.} & \textbf{\%} \\
			\hline
			Regulatory and Legal Artifacts
			& Compliance Documents and Legal Artifacts & \cite{Abualhaija2024-vj, Hassani2024-ej, Rejithkumar2024-zt, Azeem2024-az, Jain2023-aq} & 5 & 7\% \\
			\hline
			System/Software Requirements Specifications
			& System/Software Requirements Specifications & \cite{Lian2024-tk, Hasso2024-ge, Shahandashti2024-oo, Krishna2024-ck, Norheim2024-xn} & 5 & 7\% \\
			\cline{2-5}
			& Models and Designs & \cite{Ferrari2024-pu, Arora2016-nn, De_Kinderen2024-np, Ruan2023-pi} & 4 & 5\% \\
			\cline{2-5}
			& Code Artifacts & \cite{Yaacov2024-xu, Mu2024-jl, Wei2024-nc, Chang2023-zg} & 4 & 5\% \\
			\cline{2-5}
			& Requirements Analysis Reports & \cite{Fazelnia2024-hy, Lubos2024-pw, Fantechi2024-tt, Gartner2024-hw, Uygun2024-ug, Preda2024-xv} & 6 & 8\% \\
			\cline{2-5}
			& Traceability Artifacts & \cite{Hey2024-wh, North2024-td} & 2 & 3\% \\
			\hline
			Specific Quality Requirements
			& Requirements Analysis Reports & \cite{Feng2024-kj, Khakzad_Shahandashti2024-bb} & 2 & 3\% \\
			\hline
			Elicitation Documents
			& Requirements Analysis Reports & \cite{Ronanki2023-sl} & 1 & 1\% \\
			\cline{2-5}
			& Interview Scripts and Questionnaires & \cite{Gorer2024-mu} & 1 & 1\% \\
			\hline
			User Stories and Agile Requirements
			& System/Software Requirements Specifications & \cite{Li2024-zw} & 1 & 1\% \\
			\cline{2-5}
			& Models and Designs & \cite{Bragilovski2024-cv, Arulmohan2023-xf} & 2 & 3\% \\
			\cline{2-5}
			& Testing Artifacts & \cite{Kolthoff2024-dt} & 1 & 1\% \\
			\hline
			Technical Documents and Manuals
			& System/Software Requirements Specifications & \cite{Lutze2024-ao} & 1 & 1\% \\
			\cline{2-5}
			& Models and Designs & \cite{Chen2024-mj} & 1 & 1\% \\
			\cline{2-5}
			& Code Artifacts & \cite{Liao2024-my} & 1 & 1\% \\
			\cline{2-5}
			& Requirements Analysis Reports & \cite{Soares2023-xj} & 1 & 1\% \\
			\cline{2-5}
			& Testing Artifacts & \cite{Schafer2024-nf} & 1 & 1\% \\
			\hline
			Issues and User Feedback
			& System/Software Requirements Specifications & \cite{El-Hajjami2024-nc, Yang2024-ub, Tikayat_Ray2023-wm} & 3 & 4\% \\
			\cline{2-5}
			& Requirements Analysis Reports & \cite{Zhang2023-xc} & 1 & 1\% \\
			\cline{2-5}
			& User Profiles & \cite{Devathasan2024-tc, Karolita2024-rp, Wei2023-ig, Zhang2023-ml} & 4 & 5\% \\
			\hline
			Models and Ontologies
			& Process Descriptions & \cite{Klievtsova2024-ep} & 1 & 1\% \\
			\cline{2-5}
			& Models and Designs & \cite{Xie2023-wv} & 1 & 1\% \\
			\cline{2-5}
			& Code Artifacts & \cite{Nikolaidis2024-ti} & 1 & 1\% \\
			\cline{2-5}
			& Traceability Artifacts & \cite{Hassine2024-by} & 1 & 1\% \\
			\hline
			Others
			& System/Software Requirements Specifications & \cite{Binder2024-gr, Silva2024-ap} & 2 & 3\% \\
			\cline{2-5}
			& Requirements Analysis Reports & \cite{Li2024-dl, Cotroneo2024-yt} & 2 & 3\% \\
			\cline{2-5}
			& Traceability Artifacts & \cite{Rodriguez2023-wt} & 1 & 1\% \\
			\hline
		\end{tabular}%
	}
	\caption{Common input-output artifact mappings across primary studies.}
	\label{tab:common_papers_grouped_refs}
\end{table}

\begin{tcolorbox}[title=RQ2: RE Support]

\begin{itemize}
    \item Unlike NLP4RE, where \textit{Defect Detection} and \textit{Classification} were central, LLM4RE research concentrates on cognitively demanding tasks such as \textit{Elicitation} and \textit{Validation} (20\% each), as well as \textit{Modeling} (12\%) and broader \textit{Software Engineering Tasks} (15\%). This highlights the unique capacity of LLMs to support complex, human-intensive activities, and opens to greater integration with the broader software engineering field. 
    \item Compared to earlier NLP4RE surveys where \textit{Requirements Analysis} dominated, in LLM4RE studies \textit{Elicitation} \textit{and Validation \& Verification (V\&V)} are much more prominent. This shift opens automation opportunities but raises risks, since these phases involve stakeholder interaction and compliance-critical tasks.
    \item Most studies still use \textit{System/Software Requirements Specifications} as input (38\%), but there is a noticeable diversification: \textit{Issues \& User Feedback} (12\%), \textit{Regulatory \& Legal Artifacts} (9\%), as well as other less formal sources are gaining ground. This suggests LLMs make it possible to handle more complex and less structured artifact types.
    \item Beyond \textit{Specifications} (20\%) and \textit{Analysis Reports} (19\%), LLMs generate \textit{Models}, \textit{User Profiles}, \textit{Compliance Documents}, and even \textit{Code}, enabling tasks not present in earlier NLP4RE work, and showing the potential to radically expand the scope of the discipline.
    \item The field exhibits a fragmented landscape with 27 different input–output artifact pairs across 74 studies, underscoring its exploratory nature, but again confirming the versatility of LLMs.
\end{itemize}

		
		
		

\end{tcolorbox}

\subsection{ RQ3: Prompt Engineering Approaches}

In this section, we examine prompt engineering strategies (e.g., \textit{Few-shot} prompting), prompt selection strategies adopted to identify the prompts to be used in the study (e.g., \textit{Trial and Error}), and prompt engineering guidelines. The results are reported along these three dimensions in the subsequent sections.

%
%

\subsubsection{RQ3.1: Prompt Engineering Strategies}
Table~\ref{tab:combined_prompting_strategies} presents the eight broad categories of prompt engineering strategies, together with definitions and distribution. Please consider that one paper can include more than one strategy. The most frequently used approach is \textit{Zero-shot}, adopted in 28 studies (38\%), followed by  \textit{K-shot} (\textit{Few-shot} or \textit{Many-shot}), reported in 19 studies (26\%). The dominance of these simple strategies highlights a tendency to use LLMs ``out of the box'', without specific prompt engineering or tuning efforts. The focus of these studies does not appear to be on prompting, but rather on answering the question ``Can LLMs, even when used with minimal configuration, provide value in RE tasks?'' This suggests that the field is still in an exploratory phase, where demonstrating feasibility takes precedence over optimizing techniques such as Prompt Engineering, Fine-tuning, or Domain Adaptation. However, also more complex strategies are also used, often together, or in comparison, with the simpler strategies.
\textit{Reasoning and Task Decomposition} techniques are employed in 16 studies (22\%), \textit{Context-rich}, which enriches inputs with additional background or metadata, appears in 14 studies (19\%), followed by \textit{Template-based} in 13 studies (18\%). \textit{Iterative}, involving multiple rounds of refinement or interaction, is used in 10 studies (14\%). In contrast, the least frequently applied strategies are \textit{RAG-driven} (5 studies, 7\%) and \textit{Interactive} (3 studies, 4\%), suggesting that more dynamic or retrieval-augmented prompting remains underexplored in the current body of research. Overall, these results highlight a strong potential for exploring novel prompt engineering approaches in RE that move beyond the consolidated strategies currently in use. An extensive overview of prompting techniques, to be used as a map for exploration, is presented in~\cite{schulhoff2024prompt}.

\begin{table}[H]
	\centering
	\scriptsize
	\renewcommand{\arraystretch}{1.5}
	\setlength{\tabcolsep}{4pt}
	\caption{Prompt engineering strategies: definitions, references, and frequencies}
	\label{tab:combined_prompting_strategies}
    \resizebox{1\textwidth}{!}{%
	\begin{tabular}{|p{3cm}|p{6.5cm}|c|c|}
		\hline
		\textbf{Category} & \textbf{Definition / References} & \textbf{No. Studies} & \textbf{Percentage} \\
		\hline
		
		\multirow{2}{=}{Zero-shot} 
		& \textbf{Definition:} Prompting the model to perform a task without any examples or added context; relies solely on pre-trained knowledge. & \multirow{2}{*}{28} & \multirow{2}{*}{38\%} \\
		\cline{2-2}
		& \textbf{Studies:} \cite{Abualhaija2024-vj, Fazelnia2024-hy, Hassani2024-ej, Hasso2024-ge, Shahandashti2024-oo, Devathasan2024-tc, Arora2024-ib, Feng2024-kj, Hey2024-wh, Krishna2024-ck, Hassine2024-by, Khakzad_Shahandashti2024-bb, Alharbi2024-us, Sivakumar2024-jb, Gartner2024-hw, Uygun2024-ug, Kolthoff2024-dt, Karolita2024-rp, Preda2024-xv, Zhang2023-xc, El-Hajjami2024-nc, Yang2024-ub, Ruan2023-pi, Jain2023-aq, Wei2023-ig, Ronanki2023-sl, Silva2024-ap, Arora2024-gs} & & \\
		\hline
		
		\multirow{2}{=}{K-Shot: Few-shot, Many-shot}
		& \textbf{Definition:} Including a few or many examples in the prompt to guide model behavior and output style. & \multirow{2}{*}{19} & \multirow{2}{*}{26\%} \\
		\cline{2-2}
		& \textbf{Studies:} \cite{Abualhaija2024-vj, Arora2024-gs, Shahandashti2024-oo, Feng2024-kj, Krishna2024-ck, Li2024-zw, Sivakumar2024-jb, Gartner2024-hw, Kolthoff2024-dt, Karolita2024-rp, Nouri2024-jr, Rejithkumar2024-sv, Lutze2024-ao, Norheim2024-xn, Yang2024-ub, Fantechi2023-gd, Bertram2023-av, Binder2024-gr, Gorer2024-mu} & & \\
		\hline
		
		\multirow{2}{=}{Reasoning and Task Decomposition}
		& \textbf{Definition:} Step-by-step reasoning or breaking tasks into subtasks within the prompt (e.g., Chain-of-Thought). & \multirow{2}{*}{16} & \multirow{2}{*}{22\%} \\
		\cline{2-2}
		& \textbf{Studies:} \cite{Abualhaija2024-vj, Shahandashti2024-oo, Ferrari2024-pu, Feng2024-kj, Khakzad_Shahandashti2024-bb, Mu2024-jl, Alharbi2024-us, Kolthoff2024-dt, Preda2024-xv, Ruan2023-pi, Arora2016-nn, Soares2023-xj, Rodriguez2023-wt, Singhal2024-wz, Yang2023-ru, Cosler2023-uq} & & \\
		\hline
		
		\multirow{2}{=}{Context-rich}
		& \textbf{Definition:} Enriching prompts with background, metadata, or personas to better guide the model's responses. & \multirow{2}{*}{14} & \multirow{2}{*}{19\%} \\
		\cline{2-2}
		& \textbf{Studies:} \cite{Fazelnia2024-hy, Lian2024-tk, Hassani2024-ej, Hasso2024-ge, Fantechi2024-tt, Rejithkumar2024-zt, Lutze2024-ao, Cotroneo2024-yt, Silva2024-ap, Zhang2023-ml, Treude2023-er, Marczak-Czajka2023-en, Gorgen2024-ql, Li2024-dl} & & \\
		\hline
		
		\multirow{2}{=}{Template-based}
		& \textbf{Definition:} Using predefined templates to structure the model's output format or style consistently. & \multirow{2}{*}{13} & \multirow{2}{*}{18\%} \\
		\cline{2-2}
		& \textbf{Studies:} \cite{Klievtsova2024-ep, Lubos2024-pw, Yaacov2024-xu, Fantechi2024-tt, Bragilovski2024-cv, De_Kinderen2024-np, Gorgen2024-ql, Liao2024-my, Schafer2024-nf, Xie2023-wv, Veizaga2024-im, Tikayat_Ray2023-wm, Marczak-Czajka2023-ew} & & \\
		\hline
		
		\multirow{2}{=}{Iterative}
		& \textbf{Definition:} Prompting that involves multiple automatic refinement passes without human intervention. & \multirow{2}{*}{10} & \multirow{2}{*}{14\%} \\
		\cline{2-2}
		& \textbf{Studies:} \cite{Gorer2024-mu, Krishna2024-ck, North2024-td, Wei2024-nc, Ren2024-aj, Nikolaidis2024-ti, Chang2023-zg, Arulmohan2023-xf, Chen2024-mj, Chen2023-pf} & & \\
		\hline
		
		\multirow{2}{=}{RAG-driven}
		& \textbf{Definition:} Incorporating external knowledge into prompts using Retrieval-Augmented Generation (RAG) techniques. & \multirow{2}{*}{5} & \multirow{2}{*}{7\%} \\
		\cline{2-2}
		& \textbf{Studies:} \cite{Uygun2024-ug, Rejithkumar2024-sv, Azeem2024-az, Singhal2024-wz, Arora2024-gs} & & \\
		\hline
		
		\multirow{2}{=}{Interactive}
		& \textbf{Definition:} Prompts designed for back-and-forth interaction with the user (e.g., asking clarification questions). & \multirow{2}{*}{3} & \multirow{2}{*}{4\%} \\
		\cline{2-2}
		& \textbf{Studies:} \cite{Chen2024-mj, Yang2023-ru, Chen2023-pf} & & \\
		\hline
	\end{tabular}
    }
\end{table}


\begin{figure}[H]
    \centering
    \caption{Prompt engineering strategies}
    \label{fig:prompt_strategies_histogram}
    \resizebox{\textwidth}{!}{
    \begin{tikzpicture}
        \begin{axis}[
            xbar,
            width=10cm, height=5cm,
            xlabel={References (\%)},
            ylabel={Prompt Strategies},
            symbolic y coords={
                Interactive,RAG-driven,Iterative,Template-based,Context-rich,Reasoning,K-Shot,Zero-shot
            },
            ytick=data,
            yticklabel style={font=\scriptsize},
            bar width=0.4cm,
             nodes near coords={\pgfmathprintnumber{\pgfplotspointmeta}\%},
            nodes near coords style={
                /pgf/number format/.cd,
                fixed,
                fixed zerofill,
                precision=0,
                /tikz/.cd,
                font=\scriptsize
            },
            tick label style={font=\scriptsize},
            label style={font=\scriptsize},
            title style={font=\scriptsize},
            xmin=0, xmax=50
        ]
        \addplot coordinates {
            (4,Interactive) (7,RAG-driven) (14,Iterative) (18,Template-based) (19,Context-rich)
            (22,Reasoning) (26,K-Shot) (38,Zero-shot)
        };
        \end{axis}
    \end{tikzpicture}
    }
\end{figure}

\subsubsection{RQ3.2: Prompt Selection Strategies} 

Table~\ref{tab:combined_prompt_selection} presents our results concerning preliminary prompt selection strategies. The most common category is \textit{Not-specified}, observed in 29 studies (39\%), indicating that a substantial portion of the literature does not provide explicit information about how prompts were selected or refined, showing limited sensitivity to the topic, despite the high impact of prompt formulation on the output of LLMs~\cite{he2024does}.   
Among the explicitly described strategies, \textit{Trial and Error}, involving manual and systematic refinement, was used in 15 studies (20\%). \textit{Criteria-based Refinement} appears in 19 studies (26\%), showing that many studies use evaluation metrics or predefined goals to guide prompt design. Other strategies are less frequently reported. \textit{Predefined Guidelines}, which rely on domain-specific or organizational standards, were used in 6 studies (8\%). \textit{Domain Knowledge-based} approaches, incorporating subject-matter expertise, appear in 4 studies (5\%). \textit{Automatic Refinement}, which involves programmatic or AI-driven improvement of prompts, was found in only 3 studies (4\%), making it the least common approach. Overall, the data suggest that while structured strategies for prompt selection are emerging, a significant proportion of current research still relies on informal or undocumented practices. This suggests that there remains considerable room for improving LLM4RE performance, as the full potential of prompt engineering has yet to be unlocked.



\begin{table}[H]
    \centering
    \scriptsize
    \renewcommand{\arraystretch}{1.5}
    \setlength{\tabcolsep}{4pt}
    \caption{Prompt selection strategies: definitions, references, and frequencies}
    \label{tab:combined_prompt_selection}
    \resizebox{1\textwidth}{!}{%
    \begin{tabular}{|p{3.5cm}|p{5.5cm}|c|c|}
        \hline
        \textbf{Category} & \textbf{Definition / References} & \textbf{No. Studies} & \textbf{Percentage} \\
        \hline
        
        \multirow{2}{=}{Trial and Error (Manual Systematic Refinement)} 
        & \textbf{Definition:} Iterative experimentation and manual adjustment of prompts based on observations, feedback, and performance. & \multirow{2}{*}{14} & \multirow{2}{*}{19\%} \\
        \cline{2-2}
        & \textbf{Studies:} \cite{Arora2024-gs, Devathasan2024-tc, Li2024-zw, Alharbi2024-us, Uygun2024-ug, Preda2024-xv, Nouri2024-jr, Lutze2024-ao, Singhal2024-wz, Zhang2023-xc, Nikolaidis2024-ti, Gorgen2024-ql, Yang2023-ru, Rejithkumar2024-zt} & & \\
        \hline
        
        \multirow{2}{=}{Automatic Refinement (Systematic Refinement)} 
        & \textbf{Definition:} Automated methods (e.g., algorithms, feedback loops) for optimizing prompts without human intervention. & \multirow{2}{*}{2} & \multirow{2}{*}{3\%} \\
        \cline{2-2}
        & \textbf{Studies:} \cite{North2024-td, Azeem2024-az} & & \\
        \hline
        
        \multirow{2}{=}{Criteria-based Refinement} 
        & \textbf{Definition:} Refines prompts using specific criteria such as evaluation metrics or goal alignment. & \multirow{2}{*}{19} & \multirow{2}{*}{26\%} \\
        \cline{2-2}
        & \textbf{Studies:} \cite{Abualhaija2024-vj, Fazelnia2024-hy, Lubos2024-pw, Hassani2024-ej, Ferrari2024-pu, Fantechi2024-tt, Hey2024-wh, Sivakumar2024-jb, Kolthoff2024-dt, Rejithkumar2024-sv, Norheim2024-xn, De_Kinderen2024-np, El-Hajjami2024-nc, Tikayat_Ray2023-wm, Cosler2023-uq, Ruan2023-pi, Jain2023-aq, Wei2023-ig, Chen2023-pf} & & \\
        \hline
        
        \multirow{2}{=}{Predefined Guidelines} 
        & \textbf{Definition:} Refines prompts based on structured standards, templates, or documented best practices. & \multirow{2}{*}{11} & \multirow{2}{*}{15\%} \\
        \cline{2-2}
        & \textbf{Studies:} \cite{Mohedas_Daly_Loweth_Huynh_Cravens_Sienko_2022, Gorer2024-mu, arora2023llm, Krishna2024-ck, Hassine2024-by, Khakzad_Shahandashti2024-bb, Arulmohan2023-xf, Zhang2023-ml, Bragilovski2024-cv, Rodriguez2023-wt, Ekin_2023} & & \\
        \hline
        
        \multirow{2}{=}{Domain Knowledge Based} 
        & \textbf{Definition:} Refines prompts using domain-specific insights or subject-matter expertise. & \multirow{2}{*}{4} & \multirow{2}{*}{5\%} \\
        \cline{2-2}
        & \textbf{Studies:} \cite{Hasso2024-ge, Yaacov2024-xu, Gartner2024-hw, Yang2024-ub} & & \\
        \hline
        
        \multirow{2}{=}{Not Specified} 
        & \textbf{Definition:} No explicit information provided on how prompts were selected or refined. & \multirow{2}{*}{24} & \multirow{2}{*}{32\%} \\
        \cline{2-2}
        & \textbf{Studies:} \cite{Lian2024-tk, Shahandashti2024-oo, Chen2024-mj, Arora2024-ib, Feng2024-kj, Mu2024-jl, Wei2024-nc, Karolita2024-rp, Arora2016-nn, Ren2024-aj, Li2024-dl, Chang2023-zg, Treude2023-er, Marczak-Czajka2023-ew, Ronanki2023-sl, Liao2024-my, Schafer2024-nf, Xie2023-wv, Veizaga2024-im, Cotroneo2024-yt, Silva2024-ap, Fantechi2023-gd, Bertram2023-av, Soares2023-xj} & & \\
        \hline
    \end{tabular}
    }
\end{table}


\begin{figure}[H]
    \centering
    \caption{Prompt selection strategies and references}
    \label{fig:prompt_selection_histogram}
    \resizebox{\textwidth}{!}{
    \begin{tikzpicture}
        \begin{axis}[
            xbar,
            width=10cm, height=4.5cm,
            xlabel={References (\%)},
            ylabel={Prompt Strategies},
            symbolic y coords={
                Automatic Refinement, Domain Knowledge, Predefined Guidelines, Trial and Error, Criteria-based, Not Specified
                },
            ytick=data,
            yticklabel style={font=\scriptsize},
            bar width=0.4cm,
             nodes near coords={\pgfmathprintnumber{\pgfplotspointmeta}\%},
            nodes near coords style={
                /pgf/number format/.cd,
                fixed,
                fixed zerofill,
                precision=0,
                /tikz/.cd,
                font=\scriptsize
            },
            tick label style={font=\scriptsize},
            label style={font=\scriptsize},
            title style={font=\scriptsize},
            xmin=0, xmax=40
        ]
        \addplot coordinates {
            (3,Automatic Refinement) (5,Domain Knowledge) (15,Predefined Guidelines) (19,Trial and Error)
            (26,Criteria-based) (32,Not Specified)
        };
        \end{axis}
    \end{tikzpicture}
    }
\end{figure}
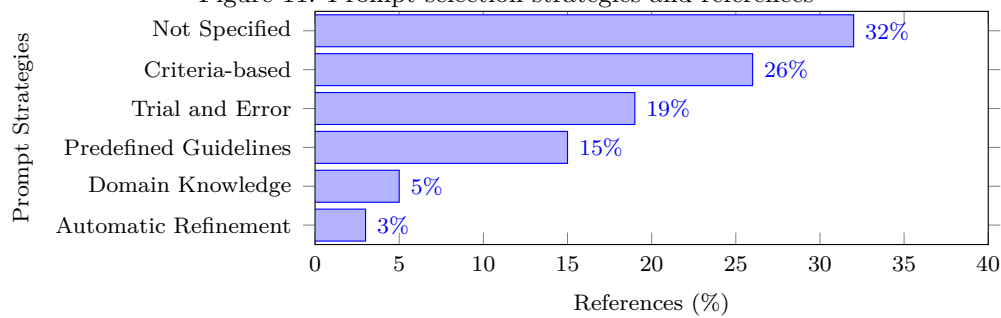

\subsubsection{RQ3.3: Prompt Engineering Guidelines}  
Table~\ref{tab:guideline_mapping} lists the guidelines along with their associated references, number of studies, and corresponding percentages out of the 74 total studies. 
The most commonly cited source is \textit{OpenAI's prompt engineering guidelines} (6 studies), followed by \textit{White et al.}~\cite{white2023promptpatterncatalogenhance} (3 studies). \textit{Mohedas et al.}~\cite{Mohedas_Daly_Loweth_Huynh_Cravens_Sienko_2022}, \textit{Arvidsson \& Axell}~\cite{arora2023llm}, \textit{Blaha \& Rumbaugh }~\cite{blaha2005object}, and \textit{Ekin et al.}~\cite{Ekin_2023} were each cited in 1 study. This indicates that only a small subset of studies (11 studies, 15\%) explicitly referenced the use of formal prompting guidelines. The results suggest a gap in standardized practices, with the majority of studies relying on informal or undocumented prompt design processes.
\begin{table}[H]
    \centering
    \scriptsize
    \caption{Prompting guidelines cited in primary studies (11 of 74 studies, 15\%)}
    \label{tab:guideline_mapping}
    \resizebox{1\textwidth}{!}{%
        \begin{tabular}{|p{6.5cm}|p{3cm}|c|c|}
            \hline
            \textbf{Guideline} & \textbf{Paper(s)} & \textbf{No. Studies} & \textbf{Percentage} \\
            \hline
            Mohedas et al. (2022)~\cite{Mohedas_Daly_Loweth_Huynh_Cravens_Sienko_2022} & \cite{Gorer2024-mu} & 1 & 1\% \\
            \hline
            Arvidsson \& Axell (2023), Arora et al. (2023)~\cite{arora2023llm} & \cite{Krishna2024-ck} & 1 & 1\% \\
            \hline
            OpenAI's prompt engineering guidelines\footnotemark & \cite{Hassine2024-by, Khakzad_Shahandashti2024-bb, Arulmohan2023-xf, Zhang2023-ml, Krishna2024-ck,Mohedas_Daly_Loweth_Huynh_Cravens_Sienko_2022,
arora2023llm,
Ekin_2023} & 8 & 11\% \\
            \hline
            Blaha \& Rumbaugh (2005)~\cite{blaha2005object} & \cite{Bragilovski2024-cv} & 1 & 1\% \\
            \hline
            White et al. (2023)~\cite{white2023promptpatterncatalogenhance} & \cite{Ferrari2024-pu, Bragilovski2024-cv, Rodriguez2023-wt} & 3 & 4\% \\
            \hline
            Ekin et al. (2023)~\cite{Ekin_2023} & \cite{Rodriguez2023-wt} & 1 & 1\% \\
            \hline
        \end{tabular}%
    }
    \footnotetext{\url{https://help.openai.com/en/articles/10032626-prompt-engineering-best-practices-for-chatgpt}}
\end{table}

\begin{tcolorbox}[title=RQ3: Prompt Engineering Approaches]
	\begin{itemize}
        \item In explored studies, prompt engineering predominantly employs simple strategies, with \textit{Zero-shot} used in 38\% of studies and \textit{K-shot} in 26\%, reflecting a focus on evaluating LLMs' feasibility for RE tasks with minimal configuration. More advanced techniques, such as \textit{Reasoning and Task Decomposition} (22\%), \textit{Context-rich} (19\%), and \textit{Template-based} (18\%), are less frequent, while \textit{RAG-driven} (7\%) and \textit{Interactive} (4\%) remain underexplored. 
        \item Prompt selection is often undocumented (39\% \textit{Not-specified}), with \textit{Trial and Error} (20\%) and \textit{Criteria-based Refinement} (26\%) being the most common  strategies. This indicates that there is still significant potential for enhancing LLM performance through more sophisticated and structured prompt engineering approaches.
		
		
	\end{itemize}
\end{tcolorbox}

\subsection{RQ4: Resources}
We now report information on the resources used in the primary studies, including datasets, available source code and prompts, and the types of LLMs adopted. 

\subsubsection{RQ4.1 Datasets} 

Table~\ref{tab:datasets} provides an overview of the datasets identified for our mapping study. In total, 10 publicly available datasets were extracted from the selected studies. These datasets were referenced by 12 out of the 74 primary studies reviewed (16\%). Each entry includes the dataset name, a brief description of its content and scope, the specific RE task(s) it supports, and the corresponding primary studies that make use of the dataset. The table highlights the diversity of datasets across various RE tasks, such as \textit{Regulatory Compliance} \cite{Abualhaija2024-vj}, \textit{Defect Detection} \cite{Fazelnia2024-hy}, \textit{Requirements Modeling} \cite{Ferrari2024-pu, Alharbi2024-us, Bragilovski2024-cv}, \textit{Requirements Validation} \cite{Lubos2024-pw, Hassine2024-by}, \textit{Requirements Elicitation} \cite{Mu2024-jl}, and \textit{Software Engineering Tasks} \cite{Li2024-zw}. For datasets associated with multiple RE tasks or studies e.g., the \textit{PURE} dataset \cite{Arora2024-ib, Ferrari2024-pu, Hasso2024-ge}, rows are subdivided to associate each RE task with its corresponding primary study. All datasets listed are publicly available and include persistent links to support accessibility and reproducibility. It should be remarked that only 16\% of the studies made use of public datasets, while many of the studies relied on private ones. This can create a replicability problem and further exacerbate the issues raised for the NLP4RE field~\cite{abualhaija2024replication}, which include difficulty in reproducing results and comparing performance across different baselines, which are the basis for any scientific discipline. 


\begin{table}[H]
    \scriptsize
	\centering
	\resizebox{1\textwidth}{!}{%
		\begin{tabular}{|p{3.5cm}|p{6cm}|p{4.5cm}|c|}
			\hline
			\textbf{Dataset} & \textbf{Description} & \textbf{RE Task} & \textbf{Ref.} \\
			\hline
			
			Annotated dataset for regulatory change\tablefootnote{\url{https://zenodo.org/records/10959496}} 
			& Data containing the annotated dataset for regulatory change analysis 
			& Legal Requirements Analysis 
			& \cite{Abualhaija2024-vj} \\
			\hline
			
			GDPR compliance decisions dataset\tablefootnote{\url{https://zenodo.org/records/11000349}} 
			& Dataset supporting classification of software requirements, specification defect identification, and conflict detection across various sources 
			& Defect Detection 
			& \cite{Fazelnia2024-hy} \\
			\hline
			
			UAV Dataset 
			& Features 400 entities across classes, properties, and individuals, with 66,296 sentences from Wikipedia and development forums 
			& Automatic text completion for software requirements 
			& \cite{Lian2024-tk} \\
			\hline
			
			\multirow{3}{*}{PURE dataset\tablefootnote{\url{https://zenodo.org/records/1414117}}} 
			& \multirow{3}{=}{Contains 79 public natural language requirements documents from the web} 
			& Standards Compliance 
			& \cite{Arora2024-ib} \\
			\cline{3-4}
			&  & Requirements Modeling 
			& \cite{Ferrari2024-pu} \\
			\cline{3-4}
			&  & Requirements Analysis 
			& \cite{Hasso2024-ge} \\
			\hline
			
			PROMISE NFR\tablefootnote{\url{https://data.mendeley.com/datasets/4ysx9fyzv4/1}} 
			& Contains 625 sentences: 255 functional and 370 non-functional requirements 
			& Requirements Validation 
			& \cite{Lubos2024-pw} \\
			\hline
			
			Rico dataset\tablefootnote{\url{https://www.kaggle.com/datasets/onurgunes1993/rico-dataset}} 
			& Mobile app UI dataset for building data-driven design applications 
			& Requirements Validation 
			& \cite{Hassine2024-by} \\
			\hline
			
			LeetCode dataset\tablefootnote{\url{https://www.kaggle.com/datasets/gzipchrist/leetcode-problem-dataset}} 
			& Contains 1,825 Leetcode problems as of April 2021 
			& Software Engineering Tasks 
			& \cite{Li2024-zw} \\
			\hline
			
			CUAD dataset\tablefootnote{\url{https://www.atticusprojectai.org/cuad}} 
			& 13,000+ labels in 510 commercial legal contracts for clause identification under legal supervision 
			& Requirements Elicitation 
			& \cite{Mu2024-jl} \\
			\hline
			
			CQ2SPARQLOWL Dataset\tablefootnote{\url{https://github.com/CQ2SPARQLOWL/Dataset}} 
			& Schema-level competency questions and their corresponding SPARQL-OWL patterns 
			& Requirements Modelling 
			& \cite{Alharbi2024-us} \\
			\hline
			
			Requirements datasets (user stories)\tablefootnote{\url{https://github.com/RELabUU/revv-light}} 
			& 22 datasets of 50+ requirements each, expressed as user stories from public or disclosed company sources 
			& Requirements Modelling 
			& \cite{Bragilovski2024-cv} \\
			\hline
			
		\end{tabular}%
	}
	\caption{Overview of publicly available datasets used in primary studies.}
	\label{tab:datasets}
\end{table}

\subsubsection{RQ4.2: Source Code and Prompts}
In total, 36 out of 74 primary studies (48\%) provided one or both artefacts, while 38 papers (52\%) did not provide supplementary materials related to their work. A comprehensive list of available source code and prompts used in these studies is provided in Appendix \ref{sec:code_availability}, specifically in Table \ref{tab:code_availability}. While this indicates a growing awareness of open science, more work is still needed to ensure replicability and verifiability of results.

\subsubsection{RQ4.3: Large Language Models}

We now explore the landscape of LLMs adopted in the studies considering the LLM types and the number of parameters. Since many studies employed different versions of LLMs from the same family (e.g., GPT-3 and GPT-4), we grouped the models according to their broader family, instead of reporting each specific model instance. In general, we identified five major families: \textit{GPT}, \textit{LLaMA}, \textit{Mixtral/Mistral}, \textit{CodeLLaMA}, and \textit{Other}; we also included \textit{BERT} as a baseline model family, as it was often used for comparison with newer LLMs---we recall that in this paper we do not classify BERT among the LLMs.  Table~\ref{tab:llm_family_summary} provides an overview of the LLM families used across the primary studies. The \textit{GPT} family is by far the most prevalent, cited in 57 of 74 studies (77\%). This is followed by \textit{LLaMA} (9 studies, 12\%), \textit{BERT} (8 studies, 11\%), and smaller families such as \textit{Mixtral/Mistral} and \textit{CodeLLaMA} (each in 3 studies, 4\%). The \textit{Other} category comprises 19 studies (26\%) and reflects the wide diversity of emerging or specialized models (e.g., Codex, GPT-J, Falcon) used in research. As several studies employed more than one model family, some references appear in multiple rows. 

Across the primary studies, the number of distinct LLM models (regardless of family) referenced per study ranged from a minimum of 1 to a maximum of 6. The median number of distinct LLMs cited per study was 1, indicating that most studies focused on a single model. Although this is in line with the exploratory nature of the studies, it should not be considered a recommended practice, as the obsolescence of models can quickly limit the validity and reproducibility of results. Evaluating multiple LLMs not only mitigates this risk but also provides a more robust understanding of how findings generalize across architectures, model families, and release cycles.
On the other hand, it is worth noting that a non-negligible number of researchers are sensitive to the problem, as 29 studies (39.19\%) utilized more than one LLM. Among these: 12 studies (16.22\%) referenced two models; 8 studies (10.81\%) cited three models; 5 studies (6.76\%) used four models; 3 studies (4.05\%) included five models; and 1 study (1.35\%) employed as many as 6 models. 

Using LLMs can have high costs if models are accessed as services, or require specialised infrastructures, and this can discourage the comparison of multiple LLMs. However, one can use strategies to mitigate this by making targeted comparisons: selecting representative models from different families, focusing on models of varying sizes or capabilities, leveraging open-source alternatives when possible, or referring to benchmark leaderboards\footnote{\url{https://huggingface.co/spaces/open-llm-leaderboard/open_llm_leaderboard}
}. Such approaches allow researchers to gain meaningful insights while controlling costs and resource requirements, without sacrificing the robustness of the study.


\begin{table}[H]
    \centering
    \scriptsize
    \caption{LLM's family usage across primary studies.}
    \renewcommand{\arraystretch}{1.5}
	\label{tab:llm_family_summary}
	\resizebox{1\textwidth}{!}{%
		\begin{tabular}{|p{2.5cm}|p{6cm}|c|c|}
			\hline
			\textbf{LLM} & \textbf{Reference} & \textbf{No.} & \textbf{\%} \\
			\hline
			GPT & \cite{Klievtsova2024-ep, Gartner2024-hw, Tikayat_Ray2023-wm, Abualhaija2024-vj, Arora2024-gs, Fantechi2024-tt, Karolita2024-rp, Preda2024-xv, Li2024-dl, De_Kinderen2024-np, Ruan2023-pi, Arulmohan2023-xf, Hassine2024-by, Li2024-zw, Alharbi2024-us, Zhang2023-xc, Ronanki2023-sl, Yang2024-ub, El-Hajjami2024-nc, Hassani2024-ej, Mu2024-jl, Rejithkumar2024-sv, Wei2023-ig, Schafer2024-nf, Silva2024-ap, Fantechi2023-gd, Xie2023-wv, Ferrari2024-pu, Chang2023-zg, Treude2023-er, Binder2024-gr, Singhal2024-wz, Wei2024-nc, Fazelnia2024-hy, Gorgen2024-ql, Arora2016-nn, Gorer2024-mu, Chen2024-mj, Kolthoff2024-dt, Nouri2024-jr, Norheim2024-xn, Ren2024-aj, Nikolaidis2024-ti, Sivakumar2024-jb, Krishna2024-ck, Chen2023-pf, Cotroneo2024-yt, Shahandashti2024-oo, Yaacov2024-xu, Khakzad_Shahandashti2024-bb, Bragilovski2024-cv, Lutze2024-ao, Marczak-Czajka2023-ew, Marczak-Czajka2023-en, Hasso2024-ge, Liao2024-my, Rodriguez2023-wt} & 57 & 77\% \\
			\hline
			LLaMA & \cite{Klievtsova2024-ep, Gartner2024-hw, Lubos2024-pw, Alharbi2024-us, Yang2024-ub, Hasso2024-ge, Rejithkumar2024-zt, Li2024-zw, Yang2023-ru} & 9 & 12\% \\
			\hline
			BERT & \cite{Lian2024-tk, Devathasan2024-tc, Hey2024-wh, Azeem2024-az, Hassani2024-ej, Tikayat_Ray2023-wm, Fazelnia2024-hy, Rejithkumar2024-zt} & 8 & 11\% \\
			\hline
			Mixtral/Mistral & \cite{Hassani2024-ej, Rejithkumar2024-sv, Hasso2024-ge} & 3 & 4\% \\
			\hline
			CodeLLaMA & \cite{North2024-td, Yang2024-ub, Krishna2024-ck} & 3 & 4\% \\
			\hline
			Other & \cite{Hey2024-wh, Jain2023-aq, Lian2024-tk, Devathasan2024-tc, Cosler2023-uq, Gorgen2024-ql, Arulmohan2023-xf, Rodriguez2023-wt, Bertram2023-av, Soares2023-xj, North2024-td, Uygun2024-ug, Wei2023-ig, Hasso2024-ge, Schafer2024-nf, Arora2016-nn, Veizaga2024-im, Fazelnia2024-hy, Arora2024-ib} & 19 & 26\% \\
			\hline
		\end{tabular}%
	}
\end{table}

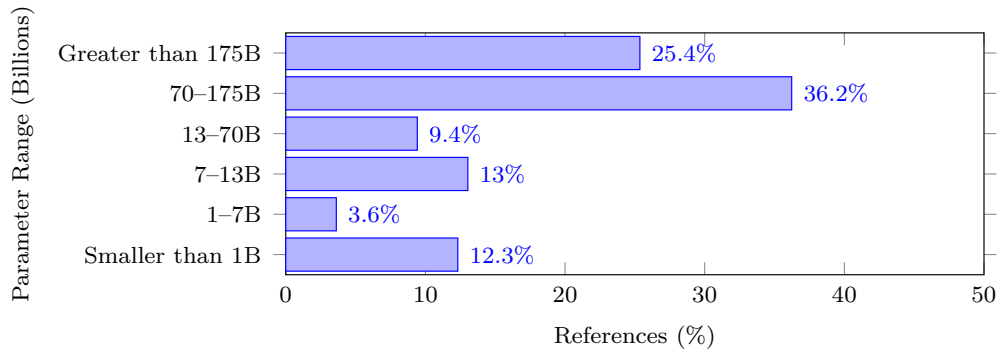
\begin{figure}[H]
    \centering
    \caption{Model mentions by parameter range}
    \label{fig:hist_model_params_ascending}
    \resizebox{\textwidth}{!}{
    \begin{tikzpicture}
        \begin{axis}[
            xbar,
            width=10cm, height=4.5cm,
            xlabel={References (\%)},
            ylabel={Parameter Range (Billions)},
            symbolic y coords={
                Smaller than 1B,1--7B,7--13B,13--70B,70--175B,Greater than 175B
            },
            ytick=data,
            yticklabel style={font=\scriptsize},
            bar width=0.4cm,
            nodes near coords={\pgfmathprintnumber{\pgfplotspointmeta}\%},
            nodes near coords style={
                /pgf/number format/.cd,
                fixed,
                precision=1,
                /tikz/.cd,
                font=\scriptsize
            },
            tick label style={font=\scriptsize},
            label style={font=\scriptsize},
            title style={font=\scriptsize},
            xmin=0, xmax=50
        ]
        \addplot coordinates {
            (12.32,Smaller than 1B) (3.62,1--7B) (13.04,7--13B) 
            (9.42,13--70B) (36.23,70--175B) (25.36,Greater than 175B)
        };
        \end{axis}
    \end{tikzpicture}
    }
\end{figure}

Fig.~\ref{fig:hist_model_params_ascending} also shows the number of parameters of the models, which indicates their size. Interestingly, many studies used LLMs with a high number of parameters, which often implies reliance on external services, since most universities and smaller organizations lack the hardware resources to run models with over 70B parameters locally---these LLMs require specialized hardware (multiple high-end GPUs or TPUs) and have huge memory requirements. While feasible for research purposes, this approach is less realistic in practical industrial settings, where companies are unlikely to expose sensitive requirements data to third-party services such as OpenAI or Google. This highlights a gap between experimental studies and real-world adoption, and suggests a potential need for smaller, locally deployable models or privacy-preserving solutions for LLM4RE applications.

\subsection{\textit{RQ4.4:} Tools and Solutions}
Table~\ref{tab:tools_combined} summarizes 21 tools and solutions identified across the literature that support a variety of RE/SE tasks, along with their corresponding references. Of these, 17  (80\%) focus on a large variety of RE tasks, from \textit{Requirements Elicitation} (5 studies, 23.8\%), to \textit{Requirements Modeling} (2 stuides, 10.3\%) and \textit{Defect Detection} (3 studies, 17.1\%), while 4 (20\%) support \textit{SE Tasks}, including code generation and
testing. These artifacts collectively represent a relatively small subset of the broader pool of primary studies. This indicates that the studies normally consider LLMs out of the box, without combining them in a meaningful way in a more complex workflows. Since most RE tasks are `hairy'~\cite{berry2021empirical}, we argue that meaningful progress will require going beyond simple, out-of-the-box applications of LLMs. Future research should investigate how LLMs can be integrated into richer workflows---possibly in combination with other AI techniques, formal methods, or domain-specific tools---to handle the complexity and multi-stakeholder nature of RE tasks.

\begin{figure}[H]
    \centering
    \scriptsize
    \caption{Distribution of solutions/tools across RE/SE tasks}
    \label{fig:tools_re_se_distribution}
    \resizebox{\textwidth}{!}{
    \begin{tikzpicture}
        \begin{axis}[
            xbar,
            width=10cm, height=8cm,
            xlabel={Tools (\%)},
            ylabel={Task},
            symbolic y coords={
               Tracing,
Terminology Extraction,
Retrieval,
Documentation,
Classification,
Change Impact Analysis,
Modeling,
Defect Detection,
SE Tasks,
Elicitation
            },
            ytick=data,
            yticklabel style={font=\scriptsize},
            bar width=0.4cm,
            nodes near coords={\pgfmathprintnumber{\pgfplotspointmeta}\%},
            nodes near coords style={
                /pgf/number format/.cd,
                fixed,
                precision=1,
                /tikz/.cd,
                font=\scriptsize
            },
            tick label style={font=\scriptsize},
            label style={font=\scriptsize},
            title style={font=\scriptsize},
            xmin=0, xmax=30
        ]
        \addplot coordinates {
            (23.8,Elicitation)
            (17.1,Defect Detection)
            (10.3,Modeling)
            (20,SE Tasks)
            (4.8,Change Impact Analysis)
            (4.8,Classification)
            (4.8,Documentation)
            (4.8,Retrieval)
            (4.8,Terminology Extraction)
            (4.8,Tracing)
        };
        \end{axis}
    \end{tikzpicture}
    }
\end{figure}

{\scriptsize
\begin{longtable}{|p{3.5cm}|p{6.5cm}|c|}
    \caption{Tools, supported RE/SE tasks, descriptions, and references} \label{tab:tools_combined} \\
    \hline
    \textbf{Tool Name} & \textbf{Task and Description} & \textbf{Reference} \\
    \hline
    \endfirsthead
    
    \multicolumn{3}{c}%
    {{\tablename\ \thetable{} Tools, supported RE/SE tasks, descriptions, and references---Continued from previous page}} \\
    \hline
    \textbf{Tool Name} & \textbf{Task and Description} & \textbf{Reference} \\
    \hline
    \endhead
    
    \hline \multicolumn{3}{|r|}{{\textit{Continued on next page}}} \\ \hline
    \endfoot
    
    \hline
    \endlastfoot
    
    MURCIA & Change Impact Analysis: Uses language models to assist in analyzing regulatory changes, supporting human analysts. & \cite{Abualhaija2024-vj} \\
    \hline
    ReqCompletion & Requirements Documentation: Recommends next token in real-time to enhance the requirements documentation process. & \cite{Lian2024-tk} \\
    \hline
    RAGTAG & SE Tasks: Generates test scenarios using Retrieval-Augmented Generation with LLMs, integrating domain knowledge. & \cite{Arora2024-gs} \\
    \hline
    GPT Agent & Requirements Elicitation: Supports eliciting requirements using a GPT-based agent. & \cite{Gorer2024-mu} \\
    \hline
    IoT Tool & Requirements Modeling: Facilitates modeling of requirements for IoT systems. & \cite{Chen2024-mj} \\
    \hline
    CGS & Requirements Terminology Extraction: Extracts Contracts Grammar to capture contractual clause essence, identifying constituents. & \cite{Rejithkumar2024-zt} \\
    \hline
    Security Tool & Requirements Tracing: Supports tracing requirements for security-related tasks. & \cite{Hassine2024-by} \\
    \hline
    ClarifyGPT & SE Tasks: Enhances code generation by identifying ambiguous requirements and asking clarifying questions using LLMs. & \cite{Mu2024-jl} \\
    \hline
    ALICE & Requirements Defect Detection: Detects contradictions in formal requirements using formal logic and LLMs. & \cite{Gartner2024-hw} \\
    \hline
    Code Gen Tool & SE Tasks: Supports software engineering tasks focused on generating code. & \cite{Wei2024-nc} \\
    \hline
    Local GPT & Requirements Retrieval: Facilitates retrieval of requirements using a localized GPT model. & \cite{Uygun2024-ug} \\
    \hline
    CRAFTER & Requirements Elicitation: Automates persona creation using GPT-3.5, recommending tailored personas. & \cite{Karolita2024-rp} \\
    \hline
    Pipeline Tool & Requirements Classification: Supports classification of requirements through a pipeline approach. & \cite{Nouri2024-jr} \\
    \hline
    nl2spec & Requirements Modeling: Derives formal specifications in temporal logics from natural language using LLMs. & \cite{Cosler2023-uq} \\
    \hline
    Mini-BAR & Requirements Elicitation: Mines user reviews in English and French using LLMs, classifying and summarizing them. & \cite{Wei2023-ig} \\
    \hline
    Weaver & Requirements Elicitation: Generates knowledge bases and recommends concepts interactively for software testing. & \cite{Yang2023-ru} \\
    \hline
    A-CodGen & SE Tasks: Generates code with fewer errors using code repository information, reducing redundancy. & \cite{Liao2024-my} \\
    \hline
    TESTPILOT & SE Tasks: Generates unit tests for JavaScript project APIs using LLMs, automating adaptive test generation. & \cite{Schafer2024-nf} \\
    \hline
    Paska & Requirements Defect Detection: Detects quality issues in natural language requirements, offering improvement recommendations. & \cite{Veizaga2024-im} \\
    \hline
    ACCA & Requirements Defect Detection: Evaluates AI-generated code correctness for security using symbolic execution against reference implementations. & \cite{Cotroneo2024-yt} \\
    \hline
    PersonaGen & Requirements Elicitation: Generates persona templates from user feedback using GPT-4 and knowledge graphs for agile analysis. & \cite{Zhang2023-ml} \\
    \hline
\end{longtable}
}

\begin{tcolorbox}[title=RQ4: Resources]
	
        \begin{itemize}
    \item Public datasets are scarce: only 10 were identified, used in 16\% of studies, with most work relying on private datasets, limiting replicability and comparability.  
    \item Source code and prompts are available in 36 out of 74 studies (48\%), showing growing awareness of open science but leaving room for improvement in transparency.  
    \item GPT-family models dominate (77\%), followed by LLaMA (12\%) and others, though almost 40\% of studies used more than one model to improve robustness.  
    \item Many studies rely on very large models (>70B parameters), which raises issues of cost, infrastructure, and privacy, and calls for smaller, locally deployable alternatives.  
    \item Only 21 tools were identified, mostly ``out-of-the-box'' uses of LLMs; future work should focus on integrating LLMs into richer workflows combining AI, formal methods, and domain-specific tools.  
\end{itemize}

\end{tcolorbox}

\subsection{RQ5: Evaluation}
We now look into the evaluation performed in the studies. In particular, we first look at how the output produced by the LLM is assessed, whether with quantitative methods, e.g., measuring precision and recall with respect to a ground truth, with qualitative one, e.g., performing thematic analysis of the output after this is generated, or with a mix of both approaches. Then, we look into the empirical evaluation strategy adopted in the study, for example, if a laboratory experiment was conducted, e.g., comparing LLMs, or if the paper reports an experimental simulation, i.e., a systematic evaluation in a less controlled, yet artificial, setting.

\subsubsection{RQ5.1: LLM Output Evaluation}
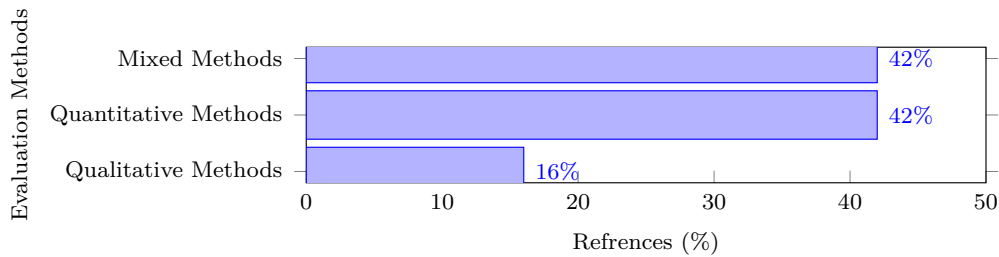
\begin{figure}[H]
    \centering
    \resizebox{\textwidth}{!}{
    \begin{tikzpicture}
        \begin{axis}[
            xbar,
            width=10cm,
            height=5cm,
            xlabel={Refrences (\%)},
            ylabel={Evaluation Methods},
            symbolic y coords={
                 Qualitative Methods,
                Quantitative Methods,
                Mixed Methods
            },
            ytick=data,
            xmin=0, xmax=50,
            bar width=0.6cm,
            y=0.7cm, 
             nodes near coords={\pgfmathprintnumber{\pgfplotspointmeta}\%},
            nodes near coords style={font=\scriptsize},
            yticklabel style={font=\scriptsize},
            tick label style={font=\scriptsize},
            label style={font=\scriptsize}
        ]
        \addplot coordinates {
            (42,Mixed Methods)
            (42,Quantitative Methods)
            (16,Qualitative Methods)
        };
        \end{axis}
    \end{tikzpicture}
    }
    \caption{Histogram of evaluation methods for LLM output and corresponding study percentages}
    \label{fig:llm_evaluation_histogram}
\end{figure}

Table\ref{tab:llm_evaluation_combined} summarizes the methods employed across primary studies for assessing LLM outputs. The methods are categorized into three types:  \textit{Quantitative Mehods} (30 studies, 42\%), which use numerical indicators for objective assessment and \textit{Qualitative Methods} (12 studies, 16\%), which rely on subjective evaluations such as expert reviews to assess, e.g.,  contextual appropriateness, completeness, or correctness of the output; \textit{Mixed Methods} (30 studies, 42\%), which combine quantitative metrics and qualitative judgments to provide an evaluation from the two different perspectives.
The distribution of evaluation methods for LLMs output across primary studies is also visualised in Fig.~\ref{fig:llm_evaluation_histogram}. \textit{Quantitative Methods} dominate together with \textit{Mixed Methods}. This suggests that LLM output cannot be evaluated only in terms of objective performance measures, but requires in-depth analysis. Indeed, LLMs are not only used to perform analytic tasks, e.g., \textit{Requirements Classification} or \textit{Defect Detection}, but are widely used for purely generative tasks, such as \textit{Model Generation}, or \textit{Requirements Elicitation}. For these  tasks, it is often hard to define a ground truth ``a priori'' to apply quantitative measures---different generated models can satisfy the same requirements, and different requirements formulations produced by an LLM can be equally adequate. Furthermore, given the richness of the output produced by an LLM, a thorough inspection is often required---possibly involving multiple judges to mitigate subjectivity---to assess, e.g., the quality of the models, or of the LLM-elicited requirements.

\begin{table}[H]
	\centering
	\scriptsize
	\renewcommand{\arraystretch}{1.5}
	\setlength{\tabcolsep}{4pt}
	\caption{Evaluation methods for LLM output: definitions, references, and frequencies}
	\label{tab:llm_evaluation_combined}
    \resizebox{1\textwidth}{!}{%
	\begin{tabular}{|p{3cm}|p{8.5cm}|c|c|}
		\hline
		\textbf{Method} & \textbf{Description and References} & \textbf{No.} & \textbf{\%} \\
		\hline
		
		Mixed Methods &
		\textbf{Definition:} Combines quantitative metrics and qualitative assessments to evaluate LLM output, balancing statistical performance with human judgment to have numerical, objective  indicators as well as in-depth insights. \newline
		\textbf{Studies:} \cite{Abualhaija2024-vj, Lian2024-tk, Arora2024-gs, Gorer2024-mu, Lubos2024-pw, Hassani2024-ej, Hasso2024-ge, Devathasan2024-tc, Yaacov2024-xu, Arora2024-ib, Feng2024-kj, Krishna2024-ck, Li2024-zw, Alharbi2024-us, Bragilovski2024-cv, Sivakumar2024-jb, North2024-td, Kolthoff2024-dt, Rejithkumar2024-sv, Azeem2024-az, Yang2024-ub, Arora2016-nn, De_Kinderen2024-np, Cosler2023-uq, Ruan2023-pi, Jain2023-aq, Marczak-Czajka2023-ew, Yang2023-ru, Xie2023-wv, Li2024-dl} &
		30 & 42\% \\
		\hline
		
		Quantitative Methods &
		\textbf{Definition:} Employs numerical measures, such as accuracy, BLEU scores, or perplexity, to objectively assess the performance and reliability of LLM-generated text. \newline
		\textbf{Studies:} \cite{Fazelnia2024-hy, Klievtsova2024-ep, Shahandashti2024-oo, Chen2024-mj, Fantechi2024-tt, Hey2024-wh, Rejithkumar2024-zt, Hassine2024-by, Khakzad_Shahandashti2024-bb, Mu2024-jl, Wei2024-nc, Uygun2024-ug, Preda2024-xv, Lutze2024-ao, Singhal2024-wz, Norheim2024-xn, Ren2024-aj, Tikayat_Ray2023-wm, Veizaga2024-im, Cotroneo2024-yt, Silva2024-ap, Zhang2023-ml, Wei2023-ig, Ronanki2023-sl, Arulmohan2023-xf, Gorgen2024-ql, Bertram2023-av, Binder2024-gr, El-Hajjami2024-nc, Rodriguez2023-wt} &
		30 & 42\% \\
		\hline
		
		Qualitative Methods &
		\textbf{Definition:} Relies on subjective analysis, such as expert reviews, to evaluate the quality, creativity, and  contextual appropriateness of LLM output. \newline
		\textbf{Studies:} \cite{Ferrari2024-pu, Gartner2024-hw, Karolita2024-rp, Nouri2024-jr, Zhang2023-xc, Nikolaidis2024-ti, Chang2023-zg, Chen2023-pf, Treude2023-er, Soares2023-xj, Marczak-Czajka2023-en, Schafer2024-nf} &
		12 & 16\% \\
		\hline
		
	\end{tabular}
    }
\end{table}

%
%

%

\subsubsection{RQ5.2: Empirical Strategy}


Table~\ref{tab:evaluation_methods_combined} summarizes the empirical strategies adopted across the primary studies, considering the classification by Stol and Fitzgerald~\cite{stol2018abc}.   The most frequently used strategy is \textit{Laboratory Experiments}, reported in 56 studies (76\%), which typically involve a systematic evaluation of the performance of LLMs against predefined ground truths. \textit{Judgment Studies}, which rely on qualitative assessments of LLM output by a small group of individuals, account for 10 studies (14\%). \textit{Field Studies}, based on observations and interviews conducted in real-world contexts without introducing interventions, appear in 5 studies (7\%). \textit{Experimental Simulations}, which attempt to replicate real-world conditions without fully controlling all the variables, are used only in 3 studies (4\%). \textit{Sample Studies} and \textit{Field Experiments} were not employed in any of the reviewed publications. 

The dominance of \textit{Laboratory Experiments} reflects the focus of the literature on controlled, repeatable evaluations of LLM performance, without considering real-world confounding factors. This is confirmed by the limited use of \textit{Field Studies} and \textit{Experimental Simulations}, which suggests a gap in understanding real-world applicability. The phenomenon is exacerbated by the absence of \textit{Sample Studies}) indicates that we currently have no evidence of the viewpoint of practitioners across different companies on LLM4RE, while the absence of \textit{Field Experiments} indicates that, while observations in specific industries have been performed (with \textit{Field Studies}), systematic evaluations are lacking. Overall, the current results suggest the pressing need to go into the field and gather evidence as well as opinions about LLM utility and effectiveness.  

\begin{table}[H]
	\centering
	\scriptsize
	\renewcommand{\arraystretch}{1.5}
	\setlength{\tabcolsep}{4pt}
	\caption{Evaluation methods used across studies: definitions, references, and statistics}
	\label{tab:evaluation_methods_combined}
     \resizebox{1\textwidth}{!}{%
	\begin{tabular}{|p{3.5cm}|p{6.5cm}|c|c|}
		\hline
		\textbf{Method} & \textbf{Description and References} & \textbf{No.} & \textbf{\%} \\
		\hline
		
		Laboratory Experiments &
		\textbf{Definition:} Rigorous evaluation of LLM against a ground truth, typically developed by authors or public datasets, without company involvement. \newline
		\textbf{Studies:} \cite{Abualhaija2024-vj, Fazelnia2024-hy, Lian2024-tk, Klievtsova2024-ep, Hassani2024-ej, Shahandashti2024-oo, Yaacov2024-xu, Chen2024-mj, Arora2024-ib, Hey2024-wh, Krishna2024-ck, Hassine2024-by, Khakzad_Shahandashti2024-bb, Mu2024-jl, Alharbi2024-us, Bragilovski2024-cv, Sivakumar2024-jb, Gartner2024-hw, North2024-td, Wei2024-nc, Uygun2024-ug, Kolthoff2024-dt, Preda2024-xv, Rejithkumar2024-sv, Lutze2024-ao, Singhal2024-wz, Azeem2024-az, Norheim2024-xn, Zhang2023-xc, Arora2016-nn, De_Kinderen2024-np, Ren2024-aj, Nikolaidis2024-ti, Li2024-dl, El-Hajjami2024-nc, Yang2024-ub, Tikayat_Ray2023-wm, Chang2023-zg, Treude2023-er, Cosler2023-uq, Ruan2023-pi, Jain2023-aq, Wei2023-ig, Arulmohan2023-xf, Chen2023-pf, Liao2024-my, Schafer2024-nf, Veizaga2024-im, Cotroneo2024-yt, Silva2024-ap, Fantechi2023-gd, Bertram2023-av, Soares2023-xj, Zhang2023-ml, Rodriguez2023-wt, Marczak-Czajka2023-en} &
		56 & 76\% \\
		\hline
		
		Experimental Simulation &
		\textbf{Definition:} Evaluation of LLM in a scenario mimicking real-world conditions, possibly compared with human subject behavior, without full control of all variables. \newline
		\textbf{Studies:} \cite{Hasso2024-ge, Fantechi2024-tt, Yang2023-ru} &
		3 & 3\% \\
		\hline
		
		Judgment Study &
		\textbf{Definition:} Manual analysis of LLM output by 2–3 individuals providing qualitative opinions. \newline
		\textbf{Studies:} \cite{Gorer2024-mu, Lubos2024-pw, Devathasan2024-tc, Ferrari2024-pu, Li2024-zw, Karolita2024-rp, Gorgen2024-ql, Binder2024-gr, Marczak-Czajka2023-ew, Ronanki2023-sl} &
		10 & 14\% \\
		\hline
		
		Sample Study &
		\textbf{Definition:} Survey-based evaluation involving many subjects, yielding quantitative results, or analysis of multiple opinions through questionnaires. \newline
		\textbf{Studies:} --- &
		0 & 0\% \\
		\hline
		
		Field Experiment &
		\textbf{Definition:} Study conducted in a real-world environment where a change is introduced, often involving company data or participants. \newline
		\textbf{Studies:} --- &
		0 & 0\% \\
		\hline
		
		Field Study &
		\textbf{Definition:} Observational study in a real-world environment involving interviews or observations without introducing a specific tool or intervention. \newline
		\textbf{Studies:} \cite{Arora2024-gs, Feng2024-kj, Rejithkumar2024-zt, Nouri2024-jr, Xie2023-wv} &
		5 & 7\% \\
		\hline
	\end{tabular}
    }
\end{table}

\begin{figure}[H]
    \centering
    \resizebox{\textwidth}{!}{
    \begin{tikzpicture}
        \begin{axis}[
            xbar,
            width=10cm,
            height=6cm,
            xlabel={Refrences (\%)},
            ylabel={Evaluation Methods},
            symbolic y coords={
                Sample Study,
                Field Experiment,
                Experimental Simulation, Field Study, Judgment Study, Laboratory Experiments
},
            ytick=data,
            xmin=0, xmax=85,
            bar width=0.6cm,
            y=0.8cm, 
             nodes near coords={\pgfmathprintnumber{\pgfplotspointmeta}\%},
            nodes near coords style={font=\scriptsize},
            yticklabel style={font=\scriptsize},
            tick label style={font=\scriptsize},
            label style={font=\scriptsize}
        ]
        \addplot coordinates {
            (76,Laboratory Experiments)
            (14,Judgment Study)
            (7,Field Study)
            (3,Experimental Simulation)
            (0,Field Experiment)
            (0,Sample Study)
        };
        \end{axis}
    \end{tikzpicture}
    }
    \caption{Histogram of evaluation methods and corresponding study percentages}
    \label{fig:evaluation_methods_histogram}
\end{figure}
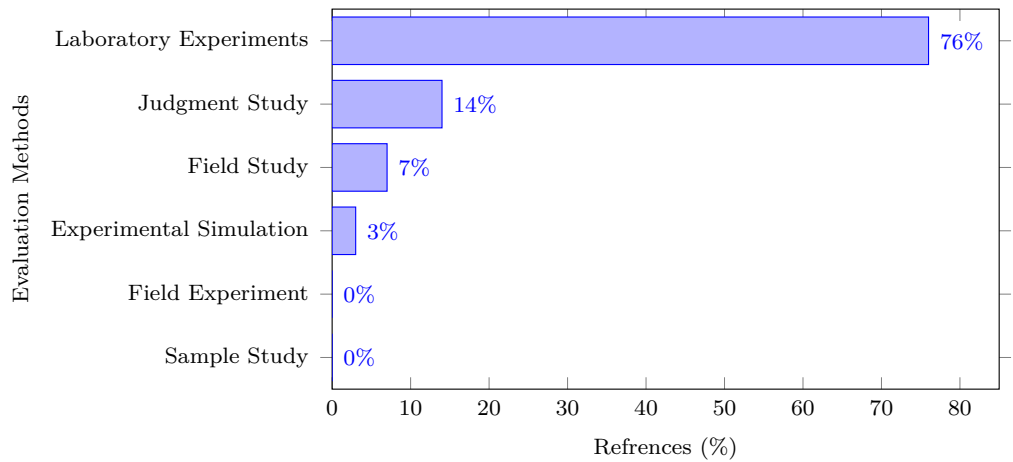

\begin{figure}[H]
    \centering
    \includegraphics[width=1\textwidth]{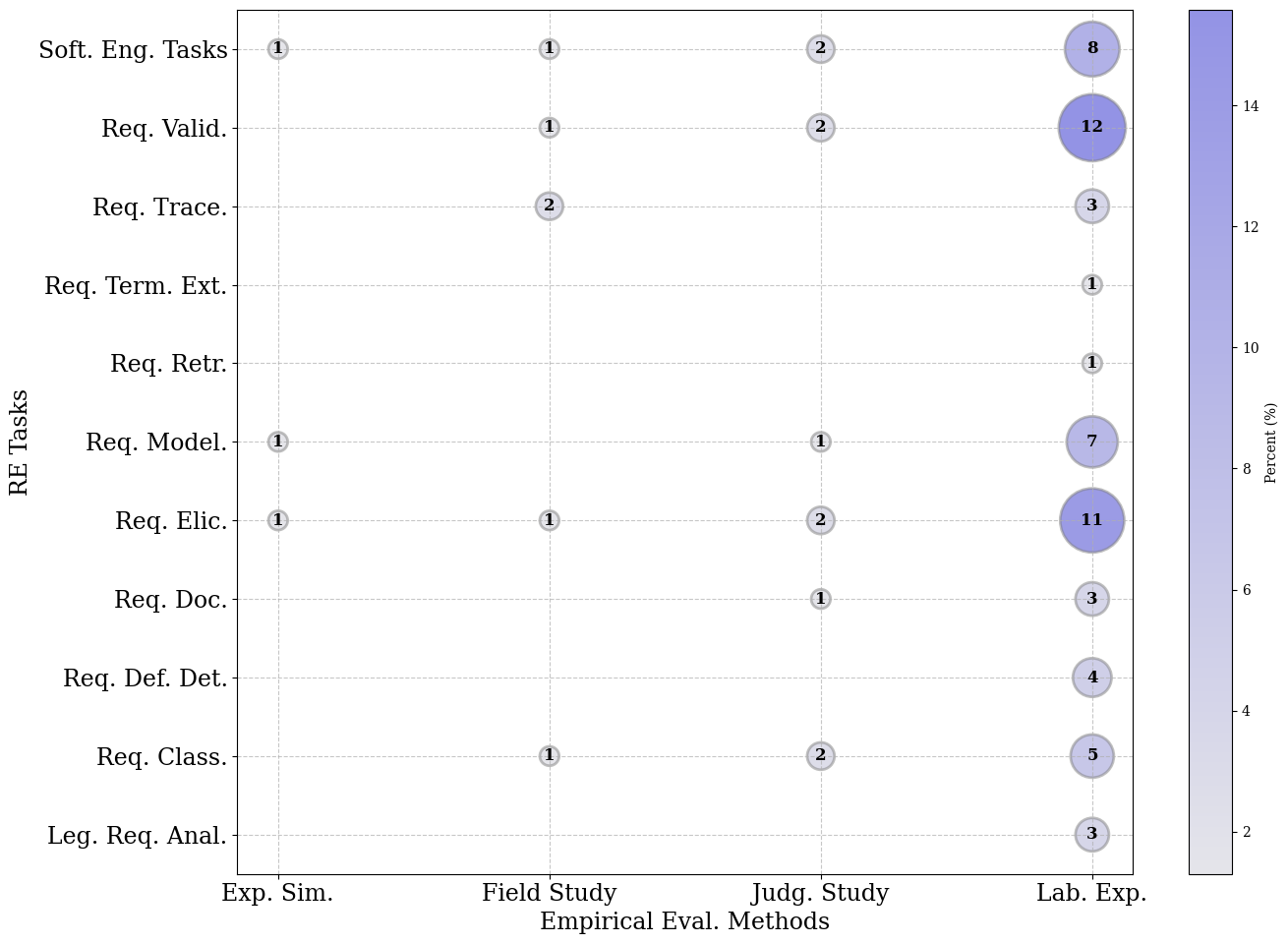}
    \caption{bubble plot of empirical strategy vs RE task}
    \label{fig:llm_emp_analysis}
\end{figure}

Figure~\ref{fig:llm_emp_analysis} analyses empirical strategies across various RE tasks. \textit{Laboratory Experiments} have been applied to all tasks, especially in \textit{Requirements Validation} (13.79\%), \textit{Requirements Elicitation} (14.94\%), \textit{Software Engineering Tasks} (12.66\%), and \textit{Requirements Modelling} (8.05\%). This diffusion highlights the flexibility of the strategy and its general suitability for all types of RE tasks. While no specific trends can be observed for the other strategies, due to their limited usage, we can observe other interesting outcomes when looking at the RE tasks: \textit{Software Engineering Tasks} and \textit{Requirements Elicitation} have been analysed with all types of empirical strategies. This suggests that the literature can provide evidence of different analysis perspectives for these tasks, considering both contrived settings~\cite{stol2018abc} (e.g., \textit{Laboratory Experiments} and \textit{Judgment Studies}), and less artificial ones (e.g., \textit{Experimental Simulations} and \textit{Field Studies}). More research adopting the different strategies is instead required for the other tasks, to demonstrate the value of LLM4RE through different empirical lenses.


\begin{tcolorbox}[title=RQ5: Evaluation]
    \begin{itemize}
    \item LLM output evaluation relies on three main approaches: \textit{Quantitative} (42\%), \textit{Qualitative} (16\%), and \textit{Mixed Methods} (42\%), reflecting the need to combine objective metrics with expert judgment, especially for those generative tasks where the output is rich, multiple correct outputs might exist, and a ground truth cannot be defined a priori (e.g., \textit{Modeling}, \textit{Elicitation}).    
    \item \textit{Laboratory Experiments} dominate empirical strategies (76\%), providing controlled and repeatable performance evaluations but limited insight into real-world applicability.  
    \item Other strategies---\textit{Judgment Studies} (14\%), \textit{Field Studies} (7\%), and \textit{Experimental Simulations} (4\%)---are underused, indicating a gap in evidence from real-world contexts and practitioner perspectives.  
    \item Certain tasks, like  \textit{Elicitation} and \textit{SE Tasks}, have been evaluated with all strategies, suggesting evidence on LLM effectiveness from different perspectives, whereas other RE tasks need more diverse empirical assessment.  
\end{itemize}
 
\end{tcolorbox}

\section{Discussion}

We now present and discuss key findings from our study, identify gaps in the existing literature, and suggest directions for future research.

\subsection{LLM4RE as a Paradigm Shift}
LLMs for RE represent a new and rapidly emerging research area, distinct from the earlier NLP4RE landscape. Unlike NLP4RE, which developed gradually over decades (1983–2019) and largely focused on analytic tasks such as defect detection and classification, LLM4RE began only in 2023, immediately following the release of ChatGPT. In just two years, 74 studies have been published, and publication counts rose 136\% in 2024 compared to 2023, suggesting that LLMs quickly captured the attention of the RE community once easy-to-use interfaces lowered technical barriers. This rapid adoption contrasts with NLP4RE, where progress was slower and heavily concentrated in specialized venues. 

In terms of tasks and artifacts, LLM4RE expands beyond the analytic focus of NLP4RE. While historical studies emphasized \textit{Requirements Defect Detection} and \textit{Classification}, LLMs are applied to cognitively demanding and generative tasks, such as \textit{Requirements Elicitation}, \textit{Validation}, and \textit{Modeling}, often leveraging diverse input sources—--from structured \textit{Requirements Specifications} to informal artifacts like \textit{User Feedback} and \textit{Issues}---and producing a broader variety of outputs. This shift demonstrates the versatility of LLMs and their capacity to tackle tasks previously considered difficult or infeasible for traditional NLP or ML approaches.

Taken together, these factors---rapid growth, broader task coverage, and  generative capabilities---suggest that LLM4RE is not merely an extension of NLP4RE, but represents a qualitatively new phase in the application of natural language analysis techniques in RE.

\subsection{LLM4RE Expands the Scope of RE}

We noted that 20\% of the studies do not strictly focus on RE tasks, but on tasks that bridge RE and SE, such as test or code generation from requirements (collectively identified as \textit{SE tasks}).  
This suggests a promising opportunity to re-center SE around requirements, and broaden the impact of the discipline. In many ways, RE becomes the foundation for other SE activities---such as coding, refactoring, testing, and maintenance---which can increasingly be automated starting from high-quality requirements. Expertise in crafting clear, complete, and unambiguous requirements is therefore crucial, not only for traditional RE tasks but also for guiding automation across the software lifecycle. In latest years, topics such as testing and mining software repositories have dominated high-profile SE venues like ICSE and journals such as TSE, while RE research has been largely absent. With the advent of LLMs, RE researchers now have a chance to re-enter these venues, demonstrating the central role of requirements in shaping RE practices.

The distribution of venues of LLM4RE studies highlights cross-disciplinary interest which goes beyond SE: whe\-re\-as NLP4RE work was largely confined to RE conferences, over 57\% of LLM4RE studies appear in diverse venues including AI, NLP, and HCI forums (e.g., HCII, LREC-COLING, MSR, CAV). This reflects the opportunity for RE researchers to engage in cross-fertilization with other domains. Building on this cross-domain potential, the field’s expertise in requirement quality can directly inform practices in prompt engineering for LLMs. Just as requirements need to be precise and interpretable to ensure correct software behavior, prompts must be carefully formulated to obtain reliable model outputs~\cite{marvin2023prompt}. In this sense, prompt engineering can be seen as a form of RE: the user specifies their needs in natural language to produce a desired result, i.e., it provides goals to be satisfied by an LLM or agent (i.e., the machine).  Consequently, RE research has the potential to expand its impact beyond its traditional community, contributing to broader AI practices, especially those where the reliability of the output is more critical, such as healthcare or other mission-critical fields.

\subsection{Rethinking LLM Applications in RE Tasks}
Despite the extended scope of tasks and artifacts considered in LLM4RE, there is still considerable room for improvement. While \textit{Classification}, \textit{Defect Detection}, and \textit{Modeling} tasks are commonly supported, the broader spectrum of tasks, such as \textit{Terminology Extraction}, \textit{Relations Extraction}, and \textit{Change Impact Analysis}, remains underrepresented. It suggests that LLMs are not yet fully leveraged for activities that require contextual reasoning and domain-specific interpretation. Furthermore, despite their increased variety, input artifacts are still strongly dominated by structured documents like \textit{Requirements Specifications} (38\%), while dynamic or informal sources such as \textit{User Feedback} (12\%) or \textit{Legal Texts} (9\%), which are increasingly critical in modern RE practice \cite{articlejacek}, are still underrepresented. Output artifacts also tend to mirror traditional deliverables, with few examples of studies generating \textit{Traceability Links},  \textit{Test Artifacts}, or \textit{Assurance Cases}. The co-occurrence of inputs/outputs highlights a lack of diversity in transformation workflows. While \textit{Requirements Specifications} are often used to generate other \textit{Specifications} (typically refined/improved), \textit{Models}, or \textit{Code}, more exploratory combinations like \textit{User Stories} to \textit{Test Cases}, or \textit{Elicitation Transcripts} to \textit{Interview Scripts} appear in only 1\% of studies each. This points to substantial opportunities for future research to expand the application of LLMs to underexplored LLM4RE tasks and input-output transformations. LLM-based approaches could greatly improve RE by supporting end-to-end workflows, especially in agile or user-centered contexts. For example, LLMs could analyze stakeholder chat transcripts to generate user stories \cite{10.1007/978-3-031-88531-0_7}, extract key terms from domain glossaries to ensure consistent terminology, or identify compliance risks by interpreting legal documents. These capabilities can help teams automate documentation, improve requirement quality, and maintain alignment with regulatory and user needs across the entire RE process. 

Another important aspect that should be considered is leveraging the generative capabilities of LLM to complement analytic tasks with \textit{explanations}. For example, tasks such as \textit{Requirements Classification} or \textit{Defect Detection} often suffer from ambiguous ground truths, as different annotators frequently interpret and label the same artifacts differently~\cite{ferrari2018detecting}. In this context, LLMs can be leveraged to provide justificatory explanations for their outputs: even when the underlying model is not inherently interpretable, it can generate convincing arguments that support and rationalize its predictions. In this sense, LLMs can not only enhance analytic tasks but also complement and improve generative and interpretative aspects of RE. Initial contributions in this direction~\cite{bashir2025requirements} have demonstrated their potential to provide justificatory explanations and increase acceptance by users.

\subsection{Toward Smarter Prompts for Requirements Engineering}

Prompt engineering in LLM-based RE approaches remains highly experimental and heterogeneous. While strategies such as \textit{Zero-shot} and  \textit{Few-shot} prompting dominate current practice, a substantial number of studies also leverage more structured techniques like task decomposition, template-based prompting, or context-rich designs. However, over 62\% of the studies combine multiple strategies, often in ad hoc ways, suggesting a lack of availability or/and knowledge about systematic frameworks for prompt composition. The limited use of interactive or RAG-based methods highlights a missed opportunity to leverage LLMs' full reasoning and retrieval capabilities for complex RE tasks such as stakeholder dialogue, traceability, or requirements validation. Prompt selection is often underreported or based on informal trial-and-error, and only a small fraction of studies explicitly reference established prompt engineering guidelines. This signals a maturity gap in methodological rigor within RE research leveraging LLMs. For practitioners, these results imply that while LLMs offer flexibility, consistent outcomes depend heavily on prompt craft quality, underscoring the need for RE-specific prompt tooling, reuse patterns, and design standards. Future work should explore the development of systematic frameworks for prompt design and evaluation, with a focus on supporting the diverse range of tasks in RE. Automating prompt construction and selection according to task type could improve consistency and scalability. In this sense, the recent study by Zadenoori et al. \cite{10.1007/978-3-031-88531-0_15}  introduced a novel approach for iteratively optimising prompts for requirements classification. This shows that there are many new creative approaches that can be studied in this field, which have not yet been examined or explored. We also recommend investigating how domain-specific ontologies and design rationale can be embedded into prompts to enhance the relevance and accuracy of LLM outputs. In addition, the RE community would benefit from empirical benchmarks that compare prompt strategies across representative RE scenarios. These benchmarks would help establish evidence-based best practices and support more informed decisions in prompt engineering.

\subsection{Openness and Reproducibility}
Openness in LLM4RE research remains insufficient. Only 12 out of 74 studies (16\%) shared publicly available datasets. Although 10 datasets were identified, they were rarely reused, which limits opportunities for benchmarking and cross-study comparison. In contrast, 45 studies (61\%) shared code or prompts, but 39\% provided no supplementary materials at all. This lack of availability restricts reproducibility and slows progress in validating results, comparing methods, and building on previous work. Without access to data, code, or prompts, researchers cannot replicate experiments, assess the strength of evidence, or evaluate how well findings generalize to other contexts. The absence of transparency also hampers the development of standardized evaluation practices across the field. Making materials such as datasets, source code, and prompts publicly available, and ensuring they are properly documented, would significantly enhance reproducibility. It would also enable meaningful cross-study validation of existing solutions. To support progress, the RE/SE community should promote shared repositories, adopt artefact availability standards, and encourage reproducible workflows. First steps in this direction have been done by the NLP4RE community itself~\cite{abualhaija2024replication}, which proposed an ID card to be associated with empirical studies to facilitate replicability. We encourage adaptation and adoption of the card also in the LLM4RE field. A framework for performing empirical studies with LLMs for SE has also been recently published~\cite{baltes2025evaluation}, with a strong focus on replicability and verifiability, giving further indications for authors to make their research truly open.

\subsection{Model Usage and Practical Deployment}
The dominance of the GPT family, cited in 77\% of the reviewed studies, reflects a strong reliance on proprietary models. Although accessible via APIs, these models raise concerns about cost, control, and reproducibility. High usage costs make it unlikely that peers will reproduce results solely to verify findings. Surprisingly, only a minority of studies explored open-source alternatives such as LLaMA or Mixtral, suggesting promising directions for future research. The common reliance on a single model per study may reflect limited resources or a lack of awareness of comparative strengths. Future work benchmarking both commercial and open-source models could yield valuable insights into their suitability for RE tasks. 

In parallel with model selection, some studies proposed practical LLM-based tools to support RE/SE tasks (21 tools in total). This is still a minority of the studies, suggesting that most of the researchers relied on LLMs ``out-of-the-box'', rather than adapting and integrating them in appropriate workflows, which could provide more robust, context-aware, and multi-step support for complex RE tasks. The predominance of ``out-of-the-box'' applications confirms that the field is still largely exploratory: researchers are testing the capabilities of LLMs on isolated tasks rather than embedding them into end-to-end processes that reflect the multi-stakeholder, iterative, and interdependent nature of RE. Future work should focus on developing integrated LLM-based workflows, combining LLMs with other AI techniques, formal methods, and domain-specific tools, to handle challenges such as traceability, validation, and requirements evolution more effectively. This shift would move LLM4RE from proof-of-concept studies toward practical, industrially relevant solutions.


\subsection{Evaluation in Practice and Novel Frameworks}
Our findings reveal that most studies (76\%) rely on \textit{Labroratory Experiments}, with limited use of methods contextualised in real-world settings like \textit{Field Studies} (7\%). While controlled settings offer consistency, they often overlook the complexity and unpredictability of actual RE practice. More \textit{Sample Studies} (e.g., surveys in industry)  need to be performed to provide evidence of the actual usage of LLM in RE in industry and the main tasks addressed, also to better target LLM4RE research towards goals that are considered relevant by practitioners. A first contribution in this sense has been given by Ullrich et al.~\cite{ullrich2025requirements}, who interviewed 18 practitioners on their use of requirements artefacts for code generation. More \textit{Field Experiments} are also needed, to avoid relying solely on practitioners' opinions, and provide evidence of LLM effectiveness in real-world scenarios. 

The tendency to use \textit{Mixed Methods}, besides \textit{Quantitative} ones, to evaluate the output of LLMs is an important shift that acknowledges the complexity of LLM output, and the need to complement numerical evaluation with in-depth qualitative ones. However, evaluation should also go beyond checking the output of LLMs, but should consider the actual utility in practice. This implies designing evaluation frameworks that measure not only correctness or similarity to a reference but also the practical impact of LLMs on RE workflows, such as time savings, error reduction, or improved stakeholder communication. By incorporating metrics for real-world \textit{effectiveness} and \textit{usability}, researchers can better assess whether LLM-based solutions deliver tangible benefits beyond lab-controlled tasks. Moreover, combining \textit{Mixed Methods} with \textit{Field Studies} and  \textit{Experimental Simulations}, also considering longitudinal observations, can provide a more holistic understanding of LLMs’ role in supporting both analytic and generative RE activities. Evaluation frameworks for LLMs such as the one proposed by Baltes \textit{et al.}~\cite{baltes2025evaluation} should be appropriately extended to include evaluation of usefulness and acceptance (e.g., with the TAM/MEM model~\cite{marangunic2015technology}), as well as other dimensions that are relevant for practice, such as the potential to reduce the cognitive workload of the users (e.g., with the NASA-TLX test~\cite{rubio2004evaluation}), while maintaining effectiveness.

\section{Threats to Validity}
In this section, we discuss the threats to validity of this study. We classified and mitigated them based on the guidelines by Apostolos \textit{et al.}~\cite{Ampatzoglou2020}.

\noindent\textbf{Study Selection Validity:} A primary threat to validity includes the formulation of the search string and its likelihood of failure in returning all the relevant papers. The threat is further increased by the use of a single database, i.e., Scopus. To address these problems, we piloted the search string and also integrated part of the consolidated string from Zhao \textit{et al.}~\cite{zhao2021natural}, which was focused on NLP4RE. Furthermore, we included a secondary search, systematically analysing a wide range of relevant RE and SE venues. This approach aimed to improve coverage and capture any relevant papers that the initial query and search engine may have missed. A further threat is the subjectivity of inclusion and exclusion criteria. To mitigate the threat, we maximised the usage of objective indicators (e.g., Core ranking, number of pages).   

\noindent\textbf{Data Validity:} The major threat to the validity of the data is the bias in data extraction due to subjectivity. This is partially mitigated by the cross-check performed by the last author---an expert in NLP4RE and LLM4RE---on a sample of the papers. This bias will be further mitigated in a future version of the paper, in which we will perform full cross-checking of the extracted data. While part of the classification schemes were derived from the literature, others were provided by the authors, which introduces subjectivity. However, the creation of the schemes was systematic and iterative, and involved two authors, which mitigates the threat. 

\noindent\textbf{Research Validity:} Research validity is guaranteed by adhering to a defined protocol, which was defined in agreement between the first and last author. All results and analyses are shared openly to enable replication and independent review. 

\section{Summary and Conclusion}
\label{sec:conclusion}
This paper presented a SLR of 74 primary studies published
between 2023 and 2024, examining the landscape of LLM4RE. 
The survey has shown that LLM4RE represents not just a continuation of NLP4RE, but a paradigm shift in how natural language technologies are applied in requirements engineering. Unlike the gradual and narrowly scoped progress of NLP4RE, the adoption of LLMs has been both rapid and expansive, catalyzed by the availability of accessible interfaces and the versatility of generative models. Within just two years, LLMs have been applied to a wide range of RE tasks, from elicitation and validation to modeling and tracing, and have begun to extend beyond RE into broader software engineering activities such as testing and code generation.

Our analysis highlights three major takeaways. First, LLM4RE re-centers requirements as a foundation for SE, offering new opportunities for RE to re-assert its central role in the discipline. Second, LLMs expand the methodological scope of RE, supporting generative and interpretative tasks, yet much potential remains untapped in underexplored areas such as terminology extraction, traceability, and compliance. Third, the success of LLM4RE depends on addressing critical gaps in openness, reproducibility, evaluation, and integration into practice. Currently, research is dominated by exploratory, proof-of-concept studies, often with limited methodological rigor, insufficient artifact sharing, and heavy reliance on proprietary models.

Looking forward, advancing LLM4RE requires both depth and breadth. Deeper progress will come from rigorous evaluation frameworks, reproducible practices, and systematic approaches to prompt engineering that recognize RE’s unique expertise in precision and clarity. Broader progress will come from expanding the scope of tasks, embracing diverse artifacts, and embedding LLMs into end-to-end RE workflows that reflect the realities of industrial practice.

In this sense, LLM4RE offers RE researchers a rare chance to reclaim centrality in SE, while simultaneously influencing the broader AI community---for example, by framing prompt engineering as a new form of requirements specification. By capitalizing on this momentum, the RE community can shape not only the future of SE, but also the reliability and trustworthiness of LLMs in safety-critical domains.



 \bibliographystyle{elsarticle-num} 
 \bibliography{cas-refs}
 \appendix
\section{Code Availability}
\label{sec:code_availability}
This appendix presents a table summarizing the code availability for RE tasks associated with specific research citations. The table includes three columns: \textbf{Citation}, \textbf{Req. Task}, and \textbf{Code}. The \textit{Citation} corresponds to the reference identifier for each study. The \textit{Req. Task} column lists the broad RE task category (e.g., Req. Elicitation, Software Engineering Tasks) derived from mapping specific tasks to a predefined set of categories based on semantic similarity. The \textit{Code} column provides hyperlinks to repositories or datasets (e.g., GitHub, Zenodo, or DOI links) where the code or data is publicly accessible; only entries with available code are included. This table serves as a resource for researchers and practitioners seeking code implementations for RE-related tasks.

\begin{longtable}{l p{0.5\linewidth} p{0.25\linewidth}}
\caption{Requirements Engineering Tasks and Code Availability} \label{tab:code_availability} \\
\toprule
\textbf{Citation} & \textbf{Req. Task} & \textbf{Code} \\
\midrule
\endfirsthead
\toprule
\textbf{Citation} & \textbf{Req. Task} & \textbf{Code} \\
\midrule
\endhead
\bottomrule
\multicolumn{3}{p{\dimexpr\linewidth-2\tabcolsep\relax}}{\small Note: Full URLs are available in the original data source. Shortened URLs are used for brevity.} \\
\endfoot
\cite{Abualhaija2024-vj} & Req. Tracing; Legal Req. Analysis & \href{https://zenodo.org/records/10890339}{Zenodo} \\
\cite{Fazelnia2024-hy} & Req. Classification; Req. Defect Detection & \href{https://zenodo.org/records/8025053}{Zenodo} \\
\cite{Klievtsova2024-ep} & Req. Modelling & \href{https://github.com/com-pot-93/m2t-trans}{GitHub} \\
\cite{Arora2024-gs} & Software Engineering Tasks & \href{https://zenodo.org/records/8049207}{Zenodo} \\
\cite{Lubos2024-pw} & Req. Validation & \href{https://github.com/MohammedAly22/GenQuest-RAG}{GitHub} \\
\cite{Hassani2024-ej} & Legal Req. Analysis & \href{https://github.com/kimixz/GPT4-Turbo-AC}{GitHub} \\
\cite{Devathasan2024-tc} & Req. Elicitation & \href{https://github.com/bThink-BGU/Papers-2024-MoDRE-BP-LLM}{GitHub} \\
\cite{Ferrari2024-pu} & Req. Modelling & \href{https://github.com/PHIKN1GHT?tab=repositories}{GitHub} \\
\cite{Arora2024-ib} & Software Engineering Tasks & \href{https://zenodo.org/record/10990762}{Zenodo} \\
\cite{Fantechi2024-tt} & Req. Elicitation & \href{https://github.com/madhava20217/LLMs-for-SRS-Prompts}{GitHub} \\
\cite{Krishna2024-ck} & Req. Terminology (Glossary) Extraction & \href{https://github.com/kimixz/GPT4-TURBO-EFFICIENCY}{GitHub} \\
\cite{Mu2024-jl} & Req. Elicitation & \href{https://github.com/SemTech23/RETROFIT-CQs}{GitHub} \\
\cite{Gartner2024-hw} & Req. Tracing & \href{https://digital.zlb.de/viewer/metadata/1364120054/}{ZLB} \\
\cite{North2024-td} & Software Engineering Tasks & \href{https://github.com/IUS-GUI-Prototyping/IUS-GUI}{GitHub} \\
\cite{Wei2024-nc} & Req. Retrieval & \href{https://doi.org/10.5281/zenodo.10149490}{DOI} \\
\cite{Uygun2024-ug} & Req. Validation & \href{https://zenodo.org/records/10649245}{Zenodo} \\
\cite{Kolthoff2024-dt} & Req. Elicitation & \href{https://zenodo.org/records/11047441}{Zenodo} \\
\cite{Preda2024-xv} & Req. Elicitation & \href{https://github.com/zhangjianzhang/ChatGPT4REIR_eval}{GitHub} \\
\cite{Singhal2024-wz} & Req. Validation & \href{https://doi.org/10.5281/zenodo.10802076}{DOI} \\
\cite{Azeem2024-az} & Req. Modelling & \href{https://github.com/KTH-RPL/DriveCmd_LLM}{GitHub} \\
\cite{Zhang2023-xc} & Req. Modelling & \href{https://doi.org/10.5281/zenodo.7745436}{DOI} \\
\cite{Nikolaidis2024-ti} & Legal Req. Analysis & \href{https://zenodo.org/record/8026646}{Zenodo} \\
\cite{Gorgen2024-ql} & Req. Validation; Legal Req. Analysis & \href{https://github.com/kimixz/GPT4-Turbo-AC}{GitHub} \\
\cite{Li2024-dl} & Software Engineering Tasks & \href{https://github.com/Jl-wei/mini-bar}{GitHub} \\
\cite{Binder2024-gr} & Req. Elicitation & \href{https://doi.org/10.5281/zenodo.8124936}{DOI} \\
\cite{El-Hajjami2024-nc} & Req. Validation & \href{https://github.com/malusamayo/Weaver}{GitHub} \\
\cite{Yang2024-ub} & Req. Documentation & \href{https://data.mendeley.com/datasets/7zbk8zsd8y/1}{Mendeley} \\
\cite{Tikayat_Ray2023-wm} & Req. Classification & \href{https://github.com/ChenKua/GRL_GPT}{GitHub} \\
\cite{Liao2024-my} & Software Engineering Tasks & \href{https://github.com/Dianshu-Liao/AAA-Code-Generation-Framework-for-Code-Repository-Local-Aware-Global-Aware-Third-Party-Aware}{GitHub} \\
\cite{Schafer2024-nf} & Software Engineering Tasks & \href{https://github.com/githubnext/testpilot}{GitHub} \\
\cite{Veizaga2024-im} & Req. Validation & \href{https://figshare.com/articles/code/Paska_-_Automated_Smell_Detection_and_Recommendation_in_Natural_Language_Requirements/22731707?file=43612056}{Figshare} \\
\cite{Cotroneo2024-yt} & Software Engineering Tasks & \href{https://github.com/dessertlab/ACCA}{GitHub} \\
\cite{Silva2024-ap} & Req. Defect Detection & \href{https://doi.org/10.5281/zenodo.13631459}{DOI} \\
\cite{Fantechi2023-gd} & Req. Defect Detection & \href{https://zenodo.org/record/8089810}{Zenodo} \\
\cite{Zhang2023-ml} & Req. Classification & \href{https://github.com/xishuozhang/PersonaGen}{GitHub} \\
\end{longtable}






\end{document}